\documentclass[11pt,onecolumn]{article}
\usepackage{setspace,dcolumn}
\usepackage{subfig}
\usepackage{hyperref}
\usepackage{cite}
\usepackage{graphicx,psfrag,epsfig,graphbox}
\usepackage[font=small,format=plain,labelfont=bf,justification=raggedright,singlelinecheck=false]{caption}
\usepackage{amsmath,amsfonts,amssymb,amsthm,bm,upgreek}
\allowdisplaybreaks[2]
\usepackage{geometry}
\usepackage[mathscr]{eucal}
\usepackage{esvect}
\usepackage{booktabs,threeparttable}
\usepackage{indentfirst}
\usepackage[normalem]{ulem}
\usepackage{lscape}
\usepackage{physics}
\usepackage{simpler-wick}
\usepackage{subfiles}
\usepackage{multirow}
\usepackage{extarrows}
\usepackage{makecell}
\usepackage{wasysym}

\usepackage{color}
\definecolor{darkgreen}{RGB}{40,150,60}
\definecolor{violet}{RGB}{140,50,230}
\definecolor{orange}{RGB}{230,150,0}
\definecolor{lightgrey}{RGB}{200,200,200}

\definecolor{redcircle}{RGB}{220,0,0}

\definecolor{greencircle}{RGB}{112,173,71}

\numberwithin{equation}{section}
\DeclareFontFamily{U}{wncy}{}
\DeclareFontShape{U}{wncy}{m}{n}{<->wncyr10}{}
\DeclareSymbolFont{mcy}{U}{wncy}{m}{n}
\DeclareMathSymbol{\Sh}{\mathord}{mcy}{"58} 

\DeclareFontEncoding{LS1}{}{}
\DeclareFontSubstitution{LS1}{stix}{m}{n}
\DeclareSymbolFont{symbols2}{LS1}{stixfrak}{m}{n}
\DeclareMathSymbol{\typecolon}{\mathbin}{symbols2}{"25}

\hypersetup{colorlinks=true,linkcolor=blue,anchorcolor=blue,citecolor=blue,urlcolor=blue}
\title{Quantization of Carrollian conformal scalar theories}
\author{Bin Chen$^{1,2,3,4}$, Haowei Sun$^3$, Yu-fan Zheng$^3$}

\begin{document}
\maketitle
\begin{center}
	{\it 
        $^{1}$Institute of Fundamental Physics and Quantum Technology, Ningbo University, Ningbo, Zhejiang 315211, China\\
        \vspace{2mm}
        $^{2}$School of Physical Science and Technology, Ningbo University, Ningbo, Zhejiang 315211, China\\
		\vspace{2mm}
		$^{3}$School of Physics, Peking University, No.5 Yiheyuan Rd, Beijing 100871, P.~R.~China\\
		\vspace{2mm}
		$^{4}$Center for High Energy Physics, Peking University, No.5 Yiheyuan Rd, Beijing 100871, P.~R.~China\\
	}
	\vspace{10mm}
\end{center}

\begin{abstract}
    \vspace{5mm}
    \begin{spacing}{1.5}
       In this work, we study the quantization of Carrollian conformal scalar theories, including two-dimensional(2D) magnetic scalar and three-dimensional(3D) electric and magnetic scalars. We discuss two different quantization schemes, depending on the choice of the vacuum. We show that the standard canonical quantization corresponding to the induced vacuum yields a unitary Hilbert space and the 2-point correlation functions in this scheme match exactly with the ones computed from the path integral. In the canonical quantization, the BMS symmetry can be realized without anomaly. On the other hand, for the quantization based on the highest-weight vacuum, it does not have a unitary Hilbert space. In 2D, the correlators in the highest-weight vacuum agree with the ones obtained by taking the $c\to 0$ limit of the 2D CFT, and there is an anomalous term in the commutation relations between the Virasoso generators, whose form is similar to the one in 2D CFT. In 3D, there is no good definition of the highest-weight vacuum without breaking the rotational symmetry. In our study, we find that the usual state-operator correspondence in CFT does not hold in the Carrollian case.  
    \end{spacing}
\end{abstract}
\newpage

\setcounter{tocdepth}{2}
\tableofcontents

\section{Introduction}

    Carrollian symmetry, the ultra-relativistic ($c \to 0$) contraction of the Poincar'e symmetry, was independently discovered by L'evy-Leblond \cite{Levy-Leblond:1965} and Sen Gupta \cite{Gupta:1966} in the 1960s. It consists of the translations along spatial the temporal directions, the rotations among spatial directions, and the Carrollian boosts defined as
    \begin{equation}
        \vec{x}~'=\vec{x}, \hspace{3ex}t'=t-\vec{b}\cdot \vec {x}.
    \end{equation}
    Subsequently, it was recognized as a viable kinematic group \cite{Bacry:1968zf}. Since then,  people have explored Carrollian symmetry and Carrollian particle dynamics in numerous works. In the $c\to 0$ limit, the light-cone collapses, resulting in trivial dynamics for a Carrollian particle which runs without moving \cite{Levy-Leblond:1965, Duval:2014uoa}. However, non-trivial dynamics may arise in complex scenarios, such as zero energy Carrollian particles \cite{deBoer:2021jej}, multiple interacting Carrollian particles \cite{Bergshoeff:2014jla, Casalbuoni:2023bbh}, one-loop effects in Carrollian scalar \cite{Banerjee:2023jpi}, Carrollian particles in electromagnetic field \cite{Marsot:2022imf}, Carrollian swifton models and Carrollian nonlinear electrodynamics \cite{Ecker:2024czx,Chen:2024vho}, and Carrollian field theories coupled to Carrollian gravity background with an extra Ehresmann connection\cite{Ciambelli:2023xqk,Ciambelli:2023tzb}. Recent progress on classical Carrollian dynamics has been covered in \cite{Marsot:2021tvq, Zhang:2023jbi} and the references therein. In recent years, Carrollian symmetry have found various applications in the study of gravitational waves \cite{Souriau:1973,Duval:2017els}, fractons \cite{Casalbuoni:2021fel, Pena-Benitez:2021ipo, Bidussi:2021nmp, Jain:2021ibh, Figueroa-OFarrill:2023vbj, Figueroa-OFarrill:2023qty, Armas:2023dcz},  gravity and cosmology\cite{Dautcourt:1997hb, Bekaert:2015xua, Hartong:2015xda, Bergshoeff:2017btm, Hansen:2021fxi, Henneaux:2021yzg, Figueroa-OFarrill:2022mcy, Bergshoeff:2023rkk, Ecker:2023uwm, deBoer:2023fnj, Tadros:2023teq}, fluid\cite{deBoer:2017ing, Ciambelli:2018xat, Ciambelli:2018wre, Petkou:2022bmz, Freidel:2022vjq}, tensionless strings \cite{Bagchi:2020fpr, Bagchi:2021rfw, Chen:2023esw}, and especially in flat holography. \par

    From a bottom-up perspective, the holographic dual of a spacetime is largely constrained by its asymptotic symmetry group (ASG). For an asymptotically flat spacetime in Einstein's gravity, its ASG was studied in the 1960s by Bondi et.al. \cite{Bondi:1962px, Sachs:1962wk, Sachs:1962zza}. This so-called Bondi-Metzner-Sachs (BMS) symmetry, including the supertranslation and superrotations,  is an extension of Poincar'e symmetry\cite{Barnich:2009se}. Quite remarkably, it turns out that the global aspect of BMS$_{d+1}$ symmetry corresponds to the $d$-dimensional Carrollian conformal symmetry \cite{Duval:2014lpa, Duval:2014uva}. As a result, the Carrollian conformal symmetry plays a key role in flat holography  \cite{Barnich:2010eb, Bagchi:2010zz, Bagchi:2012xr, Barnich:2012rz, Hartong:2015usd, Bagchi:2016bcd, Ciambelli:2019lap, Nguyen:2023vfz, Ciambelli:2018ojf,Adamo:2024mqn,Bekaert:2024itn,Have:2024dff,Chen:2023naw,Bagchi:2023fbj,Saha:2023hsl} and celestial holography \cite{Donnay:2022aba, Bagchi:2022emh, Donnay:2022wvx}.  \par

    Carrollian field theories has been an active area of study in the past few years\cite{Basu:2018dub, Barducci:2018thr, Bagchi:2019clu, Bagchi:2019xfx, Chen:2020vvn, Banerjee:2020qjj, Chen:2021xkw, Campoleoni:2021blr, Chen:2022jhx, Chen:2022cpx, Henneaux:2021yzg, Hao:2021urq, Rivera-Betancour:2022lkc, Hao:2022xhq, Yu:2022bcp, Baiguera:2022lsw, Fuentealba:2022gdx, Bergshoeff:2022eog, Bergshoeff:2022qkx, Banerjee:2022ocj, Bagchi:2022eui, Bagchi:2022eav, Bagchi:2022nvj, Saha:2022gjw, Chen:2023pqf, deBoer:2023fnj, Islam:2023rnc}.
     Among these works, the constructions of Carrollian theories were discussed in \cite{Henneaux:2021yzg} by using the Hamiltonian formalism, in \cite{Bergshoeff:2022qkx,deBoer:2023fnj} by using Galilean theories as seed theories,  and in \cite{Chen:2023pqf} by a novel method from null reduction to preserve the off-shell invariance.  In \cite{Yu:2022bcp, Hao:2021urq, Hao:2022xhq}, the authors discussed the quantization of some $2$-dimensional Carrollian theories. Additionally, there are studies on reducing massless quantum field theories in Minkowski spacetime to null infinity and doing quantization of the resulting boundary theories \cite{Liu:2022mne, Liu:2023gwa, Liu:2023qtr, Nguyen:2023vfz}. These efforts motivate further discussions on canonical quantization of Carrollian theories. \par

    In this paper, we will focus on the Carrollian conformal scalar theories in $d=2$ and $d=3$.  In $d=2$, the Carrollian conformal symmetry extends to BMS$_3$ symmetry, which contains infinite number of generators $\{l_n, m_n\}$ with $l_n$ being  Virasoro generators, leading to the super-rotations, and $m_n$ being a set of commuting generators, leading to the super-translations. In $d=3$, the Carrollian conformal symmetry extends to BMS$_4$ generated by two sets of Virasoro generators $\{l_n, \bar{l}_{\bar{n}}\}$ corresponding to super-translations, and one set of commuting generators $m_{n,\bar{n}}$ corresponding to super-translations. Since the infinite dimensional BMS$_3$ and BMS$_4$ symmetries share similar structure with the symmetry of CFT$_2$,  we can discuss them by using a similar method. In contrast,  for $d\ge 4$, the symmetry extends to BMS$_{d+1}$ symmetry, where the super-rotation part contains only finite number of generators. Thereby the Carrollian conformal theories in $d\ge 4$ should be treated in different ways.  \par

     The main motivation for this work is to  discuss the realizations of different vacua in quantum theories. The vacuum issue was initially discussed in late 80's \cite{Gamboa:1989px}. Further in \cite{Gamboa:1989zc}, the authors discussed the  quantization with Weyl ordering and the one with normal ordering, corresponding to the canonical and highest weight quantization in present work respectively. Quite recently, the vacua were discussed in \cite{Bagchi:2009pe}, and were further explored in \cite{Campoleoni:2016vsh, Casali:2016atr, Bagchi:2020fpr}. In \cite{Bagchi:2020fpr}, the authors proposed three different quantization schemes with different vacuum conditions. Here, we use  massless Carrollian scalar theories to realize the induced vacuum and the highest-weight vacuum. It turns out that in $d=2$ Carrollian magnetic scalar theory, the canonical vacuum corresponds to the induced vacuum.  We manage to find a realization of the highest-weight vacuum, which breaks the unitarity of the Hilbert space.  For $d=3$ Carrollian scalar theories, the canonical vacuum also corresponds to the induced vacuum, while there is no simple way to realize the highest-weight vacuum without breaking the rotational symmetry. \par

    Another motivation for this work is to discuss the forms of the correlation functions. In \cite{Chen:2021xkw}, the authors found two different forms of correlation functions of the Carrollian conformal field theories by solving the Ward identities, namely the power-law form and the Dirac delta-function form. In fact, both of the forms are reasonable. On one hand, the Carrollian conformal theories can be viewed as the $c\to0$ limit of CFTs, which suggests that the correlation functions should exhibit power-law behaviors of the space-time coordinates by taking the $c\to0$ limit in CFT correlators. On the other hand, it has been shown in \cite{Chen:2023pqf} that directly applying the path-integral formalism yields the correlation functions with Dirac delta-function in spatial directions multiplied by  a power-law function in the time direction, i.e. something like $t^m \partial^n\delta^{(d-1)}(x)$. In this paper, we make it clear that two forms of the correlation functions stem from different quantization schemes. More precisely, the  correlation functions in purely power-law forms correspond to the highest-weight vacuum, while the  correlation functions in delta-function forms correspond to the induced vacuum.    \par

    In an earlier paper\cite{Hao:2021urq}, the authors have valuable discussions on the highest-weight quantization of the $2$D Carrollian free electric scalar theory on the cylinder $\mathbb{R}\times S$ with non-trivial central charges. They have calculated the correlation functions, which are of power-law forms in space-time coordinates after mapping to the plane. They have further showed that this theory is not unitary in this quantization. However, the authors did not discuss the anomaly-free canonical vacuum. In fact, the discussions on the electric and magnetic scalar theories in $2$D are quite similar, and thus we do not pay much attention to the $2$D electric scalar theory in this work. Instead, we focus on the quantizations in 2D magnetic scalar and 3D electric and magnetic scalar theories. The interested readers can easily apply our discussions  to get the canonical quantization  of $2$D electric theory.  \par

    The rest of this paper is organized as follows. In section \ref{sec:2DMagneticScalar}, we examine the Carrollian magnetic scalar theory in $d=2$. We begin by providing a brief review of the BMS$_3$ symmetry, and further explore the classical symmetry of the theory. Through canonical quantization, we find that the theory behaves as infinite sets of quantum mechanics. We compute the correlation functions and find they are exactly the same as the ones computed via path integral. Next we consider alternative quantization conditions on the vacuum. It turns out that the highest-weight vacuum condition yields power-law correlation functions, although it breaks the unitarity of the Hilbert space. In section \ref{sec:3DMagneticScalar} and \ref{sec:3DElectricScalar}, we extend the discussions to $d=3$ magnetic and electric Carrollian scalar theories. The discussion of $d=3$ theories parallels to that of $d=2$ theory, except that the BMS$_4$ symmetry on $\mathbb{R}\times \mathbb{R}^2$ requires some modifications. Finally, we draw the conclusions in section \ref{sec:Discussion}. \par

\section{Magnetic scalar in \texorpdfstring{$d=2$}{d=2}}\label{sec:2DMagneticScalar}

    In this section, we discuss the quantization of massless Carrollian magnetic scalar on a cylinder $\mathbb{R}\times S$. It is natural to carry out the canonical quantization, which leads to the induced vacuum,  a BMS$_3$ invariant unitary theory, and the 2-point correlators matching with the results from path integral. It is also possible to discuss  the highest-weight vacuum, similar to the one in $2$D CFT, after giving up the unitarity. This quantization scheme leads to  correlation functions in power-law forms, which matches the structure from taking $c\to0$ limit from CFT.  \par

\subsection{BMS\texorpdfstring{$_3$}{3} symmetry}\label{subsec:BMS3Symmetry}

    Consider a theory living on a flat cylinder with coordinates $(\tau,\sigma)$, where $\tau$ is the temporal direction, and $\sigma\in[0,2\pi)$ is the spatial direction. The Carrollian manifold consists of the cylinder as the manifold, a degenerated metric $g$, and a time-like vector $\hat{\zeta}$,
    \begin{equation}\label{eq:2dCarrollianMatric}
        g = \begin{pmatrix}0&0\\0&1\end{pmatrix}, \qquad \zeta^\mu = (1, 0), \hspace{2ex}\mu=\tau,\sigma.
    \end{equation} 
    The BMS$_3$ transformations are the symmetries keeping the metric $g$ and time-like vector $\hat{\zeta}$ invariant up to an overall conformal factor, and thus the generating vectors $(\xi^\tau,\xi^\sigma)$ obey the Carrollian conformal Killing equations,
    \begin{equation}
        \partial_\tau\xi^\tau = \partial_\sigma\xi^\sigma, \quad \partial_\tau\xi^\sigma = 0.
    \end{equation}
    The solution is given by 
    \begin{equation}
        \xi^\tau = \partial_\sigma f(\sigma) \tau + g(\sigma), \qquad \xi^\sigma = f(\sigma),
    \end{equation}
    where $f(\sigma)$ and $g(\sigma)$ are arbitrary functions of $\sigma$. Thus the infinitesimal and finite coordinate transformations are respectively
    \begin{equation}\label{eq:BMS3Transformation}
        \left\{\begin{aligned} & \delta\tau = \xi^\tau = \partial_\sigma f(\sigma) \tau + g(\sigma)\\ 
            & \delta\sigma = \xi^\sigma = f(\sigma) \end{aligned}\right. ,
        \qquad \left\{\begin{aligned} & \Tilde{\tau} = \partial_\sigma F(\sigma) \tau + G(\sigma)\\ 
            & \Tilde{\sigma} = F(\sigma) \end{aligned}\right. .
    \end{equation}
    It should be stressed that the transformation on the $S^1$ part $\Tilde{\sigma} = F(\sigma)$ is an automorphism which keeps $S^1$ invariant. In fact, the corresponding algebra is the Virasoro algebra. Thus we can draw the conclusion that the BMS$_3$ transformation maps a equal-$\tau$ slice $S^1$ to another $S^1$ which may not be of equal $\tau$, and there is no BMS$_3$ transformation that can map a equal-$\tau$ slice to a point. \par
    
    The BMS$_3$ symmetry is generated by two sets of infinite number of infinitesimal transformations. The corresponding generators are
    \begin{equation}\label{eq:BMS3GeneratingVectors}
        l_n = -n\tau e^{in\sigma}\partial_\tau + ie^{in\sigma}\partial_\sigma, \quad m_n = i e^{in\sigma}\partial_\tau,\hspace{2ex} m,n\in \mathbb{Z},
    \end{equation}
    and the Lie brackets are:
    \begin{equation}
        [l_m,l_n] = (m-n) l_{m+n},\quad [l_m,m_n] = (m-n) m_{m+n},\quad [m_m,m_n] = 0.
    \end{equation}
    The corresponding finite transformations are:
    \begin{equation}
        \begin{aligned}
            &l_{n=0}: && \tilde{\tau} = \tau, && \tilde{\sigma} =\sigma + i\lambda , \\
            &l_{n\neq 0}: && \tilde{\tau} = \frac{\tau}{1+n\lambda e^{in\sigma}}, && \tilde{\sigma} =\frac{i}{n}\ln(e^{-in\sigma} + n\lambda) , \\
            & m_n: && \tilde{\tau} = \tau + i \lambda e^{in\sigma}, && \tilde{\sigma} = \sigma , \\
        \end{aligned}
    \end{equation}
    where $\lambda$ is the transformation parameter. \par

\subsection{Classical symmetries of the \texorpdfstring{$d=2$}{d=2} magnetic Carrollian scalar theory}

    In this section, we discuss the BMS$_3$ symmetry and the anisotropic scaling symmetry of the $2$D magnetic scalar theory, whose action is
    \begin{equation}\label{eq:Action2DMagneticScalar}
        S=-\frac{1}{2} \int d^2\sigma ~ 2\pi \partial_\tau \phi + \partial_\sigma \phi \partial_\sigma \phi  
    \end{equation}
    with the fundamental bosonic fields being $\pi$ and $\phi$. The equations of motion of the fundamental fields can be read from the action \eqref{eq:Action2DMagneticScalar},
    \begin{equation}
        \begin{aligned}
            \pi: & &&\partial_\tau \phi = 0, \\
            \phi: & &&\partial_\tau \pi + \partial_\sigma^2 \phi = 0.\\
        \end{aligned}
    \end{equation}
    The theory lives on the cylinder, thus we may expand the fields in terms of orthogonal and complete basis of $S^1$, i.e. the functions $e^{in\sigma}, n\in \mathrm{Z}$, 
    \begin{equation}
        \begin{aligned}
            &\phi(\tau,\sigma) = \sum_n \phi_n e^{-in\sigma} \\
            &\pi(\tau,\sigma) = \sum_n \pi_n e^{-in\sigma} + \tau n^2 \phi_n e^{-in\sigma} \\
        \end{aligned}
    \end{equation} 
    Different from the relativistic case, it is meaningless to further define the rising and lowering operators $\phi_n = a_{n} + a^\dagger_{-n}$, because $a_{n} + a^\dagger_{-n}$ always show up as a whole. \par
    
    It can be checked the theory \eqref{eq:Action2DMagneticScalar} is invariant under the BMS$_3$ transformations \eqref{eq:BMS3Transformation}. Under BMS$_3$ transformation, the fields transform as 
    \begin{equation}
        \begin{aligned} 
            &  \Tilde{\phi}(\Tilde{\tau}, \Tilde{\sigma}) =\phi(\tau, \sigma),\\
            &  \Tilde{\pi}(\Tilde{\tau}, \Tilde{\sigma}) = (\partial_\sigma F)^{-1}\left(\pi(\tau, \sigma) + \frac{\tau\partial^2_\sigma F + \partial_\sigma G}{\partial_\sigma F} \partial_\sigma\phi(\tau, \sigma) - \frac{1}{2}\left(\frac{\tau\partial^2_\sigma F + \partial_\sigma G}{\partial_\sigma F}\right)^2 \partial_\tau\phi(\tau, \sigma) \right),  \\
        \end{aligned}
    \end{equation}
    where $F$ and $G$ are the same functions in \eqref{eq:BMS3Transformation}. The infinitesimal transformations are
    \begin{equation} \label{eq:2DInfinitesimalBMSTransformation}
        \begin{aligned} 
            & \delta \phi(\tau,\sigma) = \tilde{\phi}(\Tilde{\tau}, \Tilde{\sigma}) - \phi(\tau, \sigma) = 0, \\ 
            & \delta \pi(\tau,\sigma) = \tilde{\pi}(\Tilde{\tau}, \Tilde{\sigma}) - \pi(\tau, \sigma) = -\partial_\tau\xi^\tau \pi + \partial_\sigma\xi^\tau \partial_\sigma\phi, \end{aligned}
    \end{equation}
    where $\xi$ is the vector generating the corresponding BMS$_3$ transformation. The BMS$_3$ transformations are generated by the stress tensor, which is given by
    \begin{equation}
        T^\alpha_\beta = \begin{pmatrix} -\frac{1}{2} \partial_\sigma\phi\partial_\sigma\phi & \pi\partial_\sigma\phi \\ 0 & \frac{1}{2} \partial_\sigma\phi\partial_\sigma\phi \end{pmatrix}.
    \end{equation}
    The BMS$_3$ charges can be read by $Q_\xi = -i\int d\sigma ~ \xi^\beta T^\tau_\beta$ with $\xi$ being defined in \eqref{eq:BMS3GeneratingVectors}:
    \begin{equation}\label{eq:2DBMSGeneratorsInModes}
        \begin{aligned}
            &L_n = -i\int d\sigma ~ (-n\tau T^\tau_\tau +i T^\tau_\sigma) e^{in\sigma} = \sum_{a} -2\pi i (n-a) \pi_a\phi_{n-a}, \\
            &M_n = -i \int d\sigma ~ T^\tau_\tau ~ ie^{in\sigma} = \sum_{a} \pi a(n-a) \phi_a\phi_{n-a}. \\
        \end{aligned}
    \end{equation} 
    These charges are exactly the modes in the expansion of stress tensor components
    \begin{equation}\label{eq:2DStressTensorInModes}
        \begin{aligned}
            T^\tau_\tau &= -T^\sigma_\sigma = \sum_n \frac{1}{2\pi} M_n e^{-in\sigma}, \\
            T^\tau_\sigma &= \sum_n \frac{1}{2\pi} (L_n - in\tau M_n) e^{-in\sigma}. \\
        \end{aligned}
    \end{equation}\par

\subsection{Canonical quantization} \label{subsec:2DMagneticScalarCanonicalQuantization}

    To quantize a theory means to define a Hilbert space on a equal-time slice of the space-time. The fundamental fields are the operators acting on the Hilbert space and satisfying the canonical commutation relations. In this section, we carry out the canonical quantization of the magnetic scalar theory. It turns out that the modes of the fundamental fields share similar structure with position operator $\mathbf{x}$ and momentum operator $\mathbf{p}$, which generate the Heisenberg algebras, such that the eigenstates of the Hamiltonian and hence the vacuum state are in the rigged Hilbert space. We will firstly introduce the rigged Hilbert space and then turn to the canonical quantization of the magnetic scalar theory. The canonical quantization in fact realizes the induced vacuum in the literatures.  \par

\subsubsection{Rigged Hilbert space} \label{subsec:RiggedHilbertSpace}

    In quantum mechanics, we may use a set of eigenstates $\ket{x}$ of position operator $\mathbf{x}$ as a basis of the Hilbert space  $\mathcal{H}$. Similarly, the eigenstates $\ket{p}$ of momentum $\mathbf{p}$ also make up a basis of $\mathcal{H}$. Strictly speaking, however, neither $\ket{x}$ nor $\ket{p}$ is in $\mathcal{H}$ since both of them are  Dirac delta-function normalized:
    \begin{equation}
        \bra*{x}\ket*{x^\prime} = \delta(x-x^\prime), \qquad \bra*{p}\ket*{p^\prime} = \delta(p-p^\prime).
    \end{equation}
    The mathematical structure describing such states like $\ket{x}$ or $\ket{p}$ is referred to as the rigged Hilbert space, which is also called Gelfand triplet.\cite{delaMadrid:2001dln, Madrid_2005} \par

    A generic rigged Hilbert space is defined by a triplet $(\Phi, \mathcal{H}, \Phi^\times)$ with
    \begin{equation}
        \Phi \subseteq \mathcal{H} \subseteq \Phi^\times.
    \end{equation}
    In the triplet, $\mathcal{H}$ is the Hilbert space in which the states have finite norm
    \begin{equation}
        ||\ket{\phi}|| = \sqrt{\bra{\phi}\ket{\phi}} < \infty, \qquad \forall\ket{\phi}\in\mathcal{H}, 
    \end{equation}
    $\Phi$ is the physical space in which  all the physical observables $\mathcal{O}$ have finite expectation values
    \begin{equation}
        \bra{\phi}\mathcal{O}\ket{\phi}<\infty, \qquad \forall\ket{\phi}\in\Phi,
    \end{equation}
   and $\Phi^\times$ is the dual space of $\Phi$. A state $\ket{\phi}\in\Phi$ have well-defined inner product with any state $\ket*{\phi^\times}\in\Phi^\times$:
   \begin{equation}
       (\ket{\phi}, \ket*{\phi^\times}) = \bra*{\phi^\times}\ket{\phi} <\infty, 
   \end{equation}
   where $(\ket{\phi}, \ket*{\phi^\times}) = (\ket*{\phi^\times}, \ket{\phi})^*$ denotes the inner product. Since $\ket{\phi}\in\Phi\subseteq\Phi^\times$, the norm of $\ket{\phi}$ is finite $||\ket{\phi}|| = \sqrt{\bra{\phi}\ket{\phi}}<\infty$, which coincide with the fact that $\Phi\subseteq\mathcal{H}$. Although a generic state $\ket*{\phi^\times}\in\Phi^\times$ could have divergent inner product with itself, we still formally define the inner product $\bra*{\phi^\times}\ket*{\phi^\times}$, and denote its norm by $||\ket*{\phi^\times}|| = \sqrt{\bra*{\phi^\times}\ket*{\phi^\times}}$.  \par

    For example, in the $1$D quantum mechanics, the rigged Hilbert space in the position-space representation is realized as
    \begin{equation}
        \mathcal{S}(\mathbb{R}) \subset \mathcal{H} = L^2(\mathbb{R}) \subset \mathcal{S}^\times(\mathbb{R}),
    \end{equation}
    where the Hilbert space $\mathcal{H} = L^2(\mathbb{R})$ is composed of the square integrable functions, the physical space is the Schwartz space $\mathcal{S}(\mathbb{R})$, namely the space of rapidly decreasing functions, and its dual space $\mathcal{S}^\times(\mathbb{R})$ is the space of tempered distributions. As mentioned above, the states $\ket{x}$ have divergent inner product with itself $\bra*{x}\ket*{x}\to \infty$, and thus they are not in the Hilbert space but in the dual space $\ket{x}\in\Phi^\times$. By the Gelfand-Maurin theorem, the set $\{\ket{x}\}$ serve as a basis of $\Phi$. This means for any physical state $\ket{\phi}\in\Phi$, there is a unique expansion in terms of $\ket{x}$
    \begin{equation}
        \ket{\phi} = \int dx \ket{x}\bra{x}\ket{\phi}, \qquad \forall\ket{\phi}\in\Phi,
    \end{equation}
    and the resolution of the identity is given by
    \begin{equation}
        \mathbb{I} = \int dx \ket{x}\bra{x}.
    \end{equation}
    The above discussion also applies to the states $\ket{p}$. \par
    
    It should be stressed that the states $\ket{x}$ are the ``generalized" eigenstates. Usually a state $\ket{e}\in\Phi$ is called an eigenstate of operator $\mathcal{O}$ if 
    \begin{equation}\label{eq:NormalEigenState}
        \mathcal{O}\ket{e} = e\ket{e},
    \end{equation}
    with $e$ being the eigenvalue. Consequently  in any inner product there is
    \begin{equation}
        \bra*{\phi^\times}\mathcal{O}\ket{e} = (\mathcal{O}\ket{e}, \ket*{\phi^\times}) = e (\ket{e}, \ket*{\phi^\times}) = e\bra*{\phi^\times}\ket{e}, \qquad \forall \ket*{\phi^\times}\in\Phi^\times.
    \end{equation}
    However, since $\ket{x}\in\Phi^\times$ are in the dual space, the  inner product is well defined with respect to the states in $\Phi$. Thus it is reasonable to make relation $\mathbf{x}\ket{x} = x\ket{x}$ valid only in the inner product with the states in $\Phi$:
    \begin{equation}\label{eq:GeneralizedEigenState}
        \bra{x}\mathbf{x}^\dagger\ket{\phi} = (\ket{\phi}, \mathbf{x}\ket{x}) = x (\ket{\phi}, \ket{x}) = x\bra{x}\ket{\phi}, \qquad \forall \ket{\phi}\in\Phi.
    \end{equation}
    A state $\ket{x}$ satisfying \eqref{eq:GeneralizedEigenState} is called the generalized eigenstates of $\mathbf{x}$. Notice that $\mathbf{x}\ket{x} = x\ket{x}$ does not always holds, and especially, $\mathbf{x}$ does not generically annihilate $\ket{x=0}$
    \begin{equation}
        \mathbf{x}\ket{x=0} \neq 0.
    \end{equation} 
    \par

\subsubsection{Canonical quantization and the Hilbert space}  \label{subsubsec:CanonicalQuantizationOf2DMagneticScalar}

    In flat relativistic spacetime, the Lorentzian signature differs from the Euclidean signature by the sign of the first component of the metric
    \begin{equation}
        g^{\text{L}} = \text{diag}\{-1,\vec{1}\}, \qquad g^{\text{E}} = \text{diag}\{1,\vec{1}\}.
    \end{equation}
    In the quantization, this difference causes a sign difference in the Hermitian conjugation
    \begin{equation}
        (\Phi^{\text{L}})^\dagger(\tau,\sigma)\sim\Phi^{\text{L}}(\tau,\sigma), \qquad (\Phi^{\text{E}})^\dagger(\tau,\sigma)\sim\Phi^{\text{E}}(-\tau,\sigma),
    \end{equation}
    as well as a sign difference in the canonical commutation relation:
    \begin{equation}
        [\Phi^{\text{L}},\Pi_{\Phi^{\text{L}}}]=i\delta(\sigma), \qquad [\Phi^{\text{E}},\Pi_{\Phi^{\text{E}}}]=-i\delta(\sigma),
    \end{equation}
    where $\Pi_{\Phi^{\text{A}}}$ is the canonical momentum of field $\Phi^{\text{A}}$ in signature A. However, as in \eqref{eq:2dCarrollianMatric}, the metric of the Carrollian manifold is degenerate, hence the difference between Lorentzian signature and Euclidean one in Carrollian manifold cannot be reflected through the metric. Given a Carrollian manifold, there is no prior choice of the signs in the canonical commutation relations in the quantization. In this paper, we work in the Lorentzian signature such that the Hermitian conjugation of a fundamental field $\Phi$ is $\Phi^\dagger(\tau,\sigma)\sim\Phi(\tau,\sigma)$, while the canonical quantization condition is $[\Phi,\Pi_\Phi]=i\delta(\sigma)$.\par

    By definition, the conjugation conditions of the real fields $\phi$ and $\pi$ and their modes are
    \begin{equation}
        \begin{aligned}
            &\phi^\dagger(\tau,\sigma) = \phi(\tau,\sigma), && \pi^\dagger(\tau,\sigma) = \pi(\tau,\sigma), &&\phi_n^\dagger = \phi_{-n}, && \pi_n^\dagger = \pi_{-n}. \\ 
        \end{aligned}
    \end{equation}
    The  canonical momentum of field $\phi$ is $\Pi_\phi = -\pi$. Thus the commutation relation is simply\footnote{More rigorously, it should be the Dirac comb $\Sh(\sigma) = \sum_n \delta(\sigma + 2\pi n)$ rather than Dirac delta-function $\delta(\sigma)$ since the $\sigma$ direction is periodic. }
    \begin{equation}
        [\phi(\tau, \sigma_1),-\pi(\tau, \sigma_2)]=i\delta(\sigma_1-\sigma_2), \qquad [\phi_m,\pi_n] = \frac{-i}{2\pi}\delta_{m+n}.
    \end{equation}
    Actually the micro-causality requires the commutation relation
    \begin{equation}
        \begin{aligned}\relax 
            [\phi(\tau_1, \sigma_1),\phi(\tau_2, \sigma_2)]&=0, \\
            [\phi(\tau_1, \sigma_1),\pi(\tau_2, \sigma_2)]&=-i\delta(\sigma_1-\sigma_2), \\
            [\pi(\tau_1, \sigma_1),\pi(\tau_2, \sigma_2)]&= i(\tau_1-\tau_2)\partial^2\delta(\sigma_1-\sigma_2). \\
        \end{aligned}
    \end{equation}
    The commutators vanish once $\sigma_1\neq\sigma_2$, which means the information stays at fixed spatial point. \par

    To define the Hilbert space properly, we should consider the algebra of the modes. The $\phi_n$ and $\pi_n$ modes can be divided into one set of doublet $\{\phi_0, \pi_0\}$ and infinite sets of quadruplet $\{\phi_n,\phi_{-n}, \pi_n,\pi_{-n}\}$, such that every set is the minimal one that have non-vanishing commutator and is closed under Hermitian conjugation. To make things simpler, we further divide each quadruplet set into two sets of doublet by reorganizing the modes through a Bogoliubov transformation:
    \begin{equation}
        \left\{\begin{aligned}
            \phi^c_n &= \frac{1}{\sqrt{2}}(\phi_n + \phi_{-n}),\\
            \phi^s_n &= \frac{-i}{\sqrt{2}}(\phi_n - \phi_{-n}),\\
        \end{aligned}\right. \qquad
        \left\{\begin{aligned}
            \pi^c_n &= \frac{1}{\sqrt{2}}(\pi_n + \pi_{-n}),\\
            \pi^s_n &= \frac{-i}{\sqrt{2}}(\pi_n - \pi_{-n}),\\
        \end{aligned}\right. 
        \qquad n\ge 1.
    \end{equation}
    The upper script $c$ and $s$ are for cosine and sine functions, since these modes appear in the expansion of the fields as
    \begin{equation}
        \begin{aligned}
            \phi(\tau,\sigma) &= \phi_0 + \sum_{n=1}^\infty \sqrt{2}~ \phi^c_n \cos{n\sigma} + \sqrt{2}~ \phi^s_n \sin{n\sigma} , \\
            \pi(\tau,\sigma) &= \pi_0 + \sum_{n=1}^\infty \sqrt{2} (\pi^c_n+\tau n^2 \phi^c_n) \cos{n\sigma} + \sqrt{2} (\pi^s_n+\tau n^2 \phi^s_n) \sin{n\sigma}, \\
        \end{aligned}
    \end{equation}
    and moreover, all the modes are Hermitian now
    \begin{equation}
        \begin{aligned}
            &(\phi_0)^\dagger = \phi_0, \quad(\phi^c_n)^\dagger = \phi^c_n, \quad (\phi^s_n)^\dagger = \phi^s_n, \\
            &(\pi_0)^\dagger = \pi_0, \quad (\pi^c_n)^\dagger = \pi^c_n, \quad (\pi^s_n)^\dagger = \pi^s_n. \\
        \end{aligned}
    \end{equation}
    Thus the modes in the theory are divided into the sets,
    \begin{equation} \label{eq:2DSetsOfHeisenbergAlgebra}
        \{\phi_0,\pi_0\}, \quad \{\phi^c_n,\pi^c_n\}, \quad \{\phi^s_n,\pi^s_n\},
    \end{equation}
    each of which makes up a Heisenberg algebra since the commutation relations between the modes in \eqref{eq:2DSetsOfHeisenbergAlgebra} are
    \begin{equation}
        \begin{aligned}\relax 
            [\phi_0,\pi_0] & = \frac{-i}{2\pi}, &
            [\phi^c_m,\pi^c_n] & = \frac{-i}{2\pi}\delta_{m,n}, &
            [\phi^s_m,\pi^s_n] & = \frac{-i}{2\pi}\delta_{m,n}.
        \end{aligned}
    \end{equation} 
    This result implies that the theory is  made up of infinite sets of quantum mechanics \eqref{eq:2DSetsOfHeisenbergAlgebra}. \par
    
    The Hamiltonian $H = -M_0$ in terms of these modes
    \begin{equation}
        \begin{aligned}
            H &= \sum_{a=-\infty}^\infty \pi a^2 \phi_a\phi_{-a} = \sum_{a=-\infty}^\infty \pi a^2 \abs{\phi_a}^2 = \sum_{a=1}^\infty \pi a^2((\phi^c_a)^2 + (\phi^s_a)^2)
        \end{aligned}
    \end{equation}
    is bounded below. The energy eigenstate is also the eigenstates of $\phi^{c/s}_n$ modes, thus are similar to the position eigenstate in quantum mechanics\footnote{One may consider the Fock space structure with creating and annihilating operator $a^{c/s}_n \sim \phi^{c/s}_n + i\pi^{c/s}_n, a^{c/s}_n \sim \phi^{c/s}_n - i\pi^{c/s}_n$. However, these states are not eigenstates of the Hamiltonian. }.
    Noticing that there is no preference state under $\phi_0$ and $\pi_0$ modes, since these modes are absent in the Hamiltonian. In this work, we choose the basis states to be the eigenstates of $\phi_0$ such that the correlation functions from canonical quantization agree with the ones from path integral quantization. In summary, the eigenstates are the tensor product of
    \begin{equation}
        \ket{\alpha} = \ket{\alpha_0}\otimes\ket{\alpha^c_{1}}\otimes\ket{\alpha^s_{1}}\otimes\ket{\alpha^c_{2}}\otimes\ket{\alpha^s_{2}}\otimes\cdots ,\label{eigenstate}
    \end{equation}
    with
    \begin{equation}
        \begin{aligned}
            &\phi_0\ket{\alpha} = \alpha_0 \ket{\alpha}, \qquad \phi^{c/s}_n \ket{\alpha} = \alpha^{c/s}_n \ket{\alpha},  \\
            &\phi_n \ket{\alpha} = \frac{1}{\sqrt{2}}(\alpha^c_n + i \alpha^s_n) \ket{\alpha} = \alpha_n\ket{\alpha}, \\
            &\phi_{-n} \ket{\alpha} = \frac{1}{\sqrt{2}}(\alpha^c_n - i \alpha^s_n) \ket{\alpha} = \alpha_{-n}\ket{\alpha}.\\
        \end{aligned}
    \end{equation}
    The canonical vacuum is chosen to be 
    \begin{equation}
        \ket{\text{vac}} =\ket{\alpha = 0} .
    \end{equation} 
    This vacuum is the lowest energy state $\bra{\text{vac}}H\ket{\text{vac}}=0$ and has vanishing vacuum expectation values(VEVs) for all the BMS$_3$ generators, which will be discussed shortly. This canonical vacuum corresponds to the induced vacuum in \cite{Bagchi:2020fpr}. One disadvantage of the above choice is that the eigenstate \eqref{eigenstate} and the vacuum are non-normalizable. Indeed, it follows the general discussions in quantum mechanics that 
    \begin{equation}
        \bra{\alpha} = \ket{\alpha}^\dagger = \bra{\alpha_0}\otimes\bra{\alpha^c_{1}}\otimes\bra{\alpha^s_{1}}\otimes\bra{\alpha^c_{2}}\otimes\bra{\alpha^s_{2}}\otimes\cdots ,
    \end{equation}
    and thus the inner products is defined as 
    \begin{equation}
        \braket{\tilde{\alpha}}{\alpha} = \braket{\tilde{\alpha}_0}{\alpha_0} \prod_{n=1}^\infty \braket{\tilde{\alpha}^c_n}{\alpha^c_n} \prod_{n=1}^\infty \braket{\tilde{\alpha}^c_n}{\alpha^c_n} = \delta(\alpha_0 -\tilde{\alpha}_0) \prod_{n=1}^\infty \delta(\alpha^c_0 -\tilde{\alpha}^c_0) \prod_{n=1}^\infty \delta(\alpha^s_0 -\tilde{\alpha}^s_0).
    \end{equation}
    The inner product $\braket{\tilde{\alpha}}{\alpha}$ is divergent as $\tilde{\alpha}\to\alpha$, and the non-normalizable nature of the states leads to  divergence in $\pi\pi$ propagator. \par

    Nevertheless, we can consider the expectation value $\expval{\mathcal{O}}_{\alpha}$ of operator $\mathcal{O}$ in the state $\ket{\alpha}$ which is defined as 
    \begin{equation}
        \expval{\mathcal{O}}_{\alpha} = \frac{\bra{\alpha}\mathcal{O}\ket{\alpha}}{\bra{\alpha}\ket{\alpha}} = \lim_{\tilde{\alpha}\rightarrow\alpha}\frac{\bra{\tilde{\alpha}}\mathcal{O}\ket{\alpha} + \bra{\alpha}\mathcal{O}\ket{\tilde{\alpha}}}{2\bra{\tilde{\alpha}}\ket{\alpha}} ,
    \end{equation}
    where the subscript $\alpha$ is for the state $\ket{\alpha}$. This definition is compatible with Hermiticity of the operator $\mathcal{O}$:
    \begin{equation}
        \expval*{\mathcal{O}^\dagger}_{\alpha} = \lim_{\tilde{\alpha}\rightarrow\alpha}\frac{\bra{\tilde{\alpha}}\mathcal{O}^\dagger\ket{\alpha} + \bra{\alpha}\mathcal{O}^\dagger\ket{\tilde{\alpha}}}{2\bra{\tilde{\alpha}}\ket{\alpha}} = \lim_{\tilde{\alpha}\rightarrow\alpha}\frac{\bra{\alpha}\mathcal{O}^\dagger\ket{\tilde{\alpha}}^* + \bra{\tilde{\alpha}}\mathcal{O}^\dagger\ket{\alpha}^* }{2\bra{\tilde{\alpha}}\ket{\alpha}} = \expval*{\mathcal{O}}_{\alpha}^*
    \end{equation}
    where we used $\bra{\tilde{\alpha}}\ket{\alpha}^* = \bra{\tilde{\alpha}}\ket{\alpha}$. Thus a Hermitian operator has  real expactation values. \par

    Similar to the discussions in quantum mechanics, we can also define the eigenstates $\ket{\kappa} = \ket{\kappa_0}\otimes\ket{\kappa^c_{1}}\otimes\ket{\kappa^s_{1}}\otimes\ket{\kappa^c_{2}}\otimes\ket{\kappa^s_{2}}\otimes\cdots$ of $\pi_0$ and $\pi^{c/s}_n$ modes, and find the overlapping with $\ket{\alpha}$
    \begin{equation}
        \begin{aligned}
            \braket{\alpha}{\kappa} = \braket{\kappa}{\alpha}^* & = \exp{-2\pi i\kappa_0\alpha_0}\prod_{n=1}^\infty \exp{-2\pi i(\kappa^c_n\alpha^c_n + \kappa^s_n\alpha^s_n)} \\
            & = \prod_{n\in\mathbb{Z}} \exp{-2\pi i\kappa_{n}\alpha_{-n}} = \exp{-2\pi i\sum_{n}\kappa_{n}\alpha_{-n}}. \\
        \end{aligned}
    \end{equation}
    Thus we can insert the identity operator $\mathbb{I} = \int (\prod d\kappa) \ketbra{\kappa}$ or $\mathbb{I} = \int (\prod d\alpha) \ketbra{\alpha}$ into the correlators to simplify the  calculations. \par

    It should be stressed that the only meaningful quantity is the VEVs of the operators. As discussed in \ref{subsec:RiggedHilbertSpace}, the state $\phi_0\ket{\text{vac}}$ is not always null.  For example, it will be proved in \eqref{eq:2dVEVofPhi0Pi0} that the inner product of $\pi_0\ket{\text{vac}}$ and $\phi_0\ket{\text{vac}}$ is non-vanishing,
    \begin{equation}
        \bra{\text{vac}}\pi_0\phi_0\ket{\text{vac}} = \frac{-i}{4\pi} \neq 0.
    \end{equation}
    Thus, it is only reasonable to say $\expval{\phi_0}=0$, while $\phi_0\ket{\text{vac}}$ is not strictly null.  \par
    
    Now that the quantization is defined, let us discuss the symmetry generators. By definition, the VEVs of most of the generators of the BMS$_3$ symmetries and the anisotropic symmetries are vanishing
    \begin{equation}
        \expval{L_{n\neq 0}} = \expval{M_{n}} = 0.
    \end{equation}
    The operator $L_0$ has ambiguity in  the ordering of the modes \eqref{eq:2DBMSGeneratorsInModes}. We choose the Weyl ordering such that the generator is Hermitian,
    \begin{equation}
        \begin{aligned}
            L_0 &= \sum_{a} \pi i a(\pi_a\phi_{-a} + \phi_{-a}\pi_a),
        \end{aligned}
    \end{equation}
    with $(L_0)^\dagger = L_0$. In this ordering, its VEV is vanishing, $\expval{L_0}=0$. Considering the VEVs of  commutators $[L_n,L_m]$, we see that the anomaly term $a_L(m)$ in $[L_m,L_n] = (m-n)L_{m+n} + a_L(m)\delta_{m+n} $ vanishes, 
    \begin{equation}
        \begin{aligned}
            0=\expval{[L_n,L_{-n}]} = 2n \expval{L_0} + a_L(n) = a_L(n).
        \end{aligned}
    \end{equation}
    There is no ordering ambiguity in $M_0$, and the anomaly $a_M(m)$ in $[L_m, M_n] = (m-n)M_{m+n} + a_M(m)\delta_{m+n}$ vanishes as well. Thus the theory is anomaly free. In this ordering, the tress tensor is Hermitian $(T^\mu_\nu)^\dagger = T^\mu_\nu$ by \eqref{eq:2DStressTensorInModes}. In the vacuum, the expectation value of the stress tensor vanishes
    \begin{equation}
        \expval{T^\mu_\nu(\tau,\sigma)} = 0.
    \end{equation}
    Physically this means that the vacuum contains no energy, momentum and stress. \par

    In conclusion, with respect to the vacuum, the generators of the symmetry algebra obey the following commutation relations
    \begin{equation}
        \begin{aligned}
            &[L_m, L_n] = (m-n)L_{m+n}, \qquad [L_m, M_n] = (m-n)M_{m+n}, \qquad [M_m, M_n] = 0.
        \end{aligned}
    \end{equation}
    These relations can be recovered by the realization of the generators in term of the modes. Besides, the action of a generator $G$ on the field $\Phi$ gives  the corresponding infinitesimal transformation $\delta \Phi = \xi^\mu_G\partial_\mu \Phi + [G, \Phi]$. It can be checked that the actions indeed match with \eqref{eq:2DInfinitesimalBMSTransformation}:
    \begin{align}
        &\begin{aligned}
            & \xi^\tau_{L_n} = n\tau e^{in\sigma}, \hspace{3ex}\xi^\sigma_{L_n} = -i e^{in\sigma}, \\
            & [L_n, \phi] = i e^{in\sigma}\partial_\sigma\phi = -\xi^\mu_{L_n}\partial_\mu\phi, \\
            & [L_n, \pi] =  -n \tau e^{in\sigma} \partial_\tau \pi + i e^{in\sigma}\partial_\sigma\pi - n e^{in\sigma} \pi + in^2 \tau e^{in\sigma} \partial_\sigma\phi\\
                & \qquad\quad = -\xi^\mu_{L_n}\partial_\mu\pi -\partial_\tau\xi^\tau_{L_n} \pi + \partial_\sigma\xi^\tau_{L_n} \partial_\sigma\phi,\\
        \end{aligned}\\
        &\begin{aligned}
            & \xi^\tau_{M_n} = -i e^{in\sigma},\hspace{3ex} \xi^\sigma_{M_n} = 0, \\
            &[M_n, \phi] = 0 = -\xi^\mu_{M_n}\partial_\mu\phi,\\
            & [M_n, \pi] = i e^{in\sigma} \partial_\tau \pi + n e^{in\sigma} \partial_\sigma\phi = -\xi^\mu_{M_n}\partial_\mu\pi -\partial_\tau\xi^\tau_{M_n} \pi + \partial_\sigma\xi^\tau_{M_n} \partial_\sigma\phi, \\
        \end{aligned}
    \end{align}

    By using the mode expansions, we can calculate the 2-point correlation functions.  It is easy to see that VEVs of two $\phi_n$ modes are always vanishing 
    \begin{equation}
        \expval{\phi_m\phi_n} = 0.
    \end{equation}
    For the correlator between one $\phi_n$ and one $\pi_n$, we should insert the identity to do the calculation. For example, we have
    \begin{equation}\label{eq:2dVEVofPhi0Pi0}
        \begin{aligned}
            &\expval{\phi_0\pi_0} 
                = \lim_{\alpha\rightarrow 0}\lim_{\tilde{\alpha}\rightarrow\alpha}\frac{1}{2\bra{\tilde{\alpha}}\ket{\alpha}} \left(\bra{\tilde{\alpha}}\phi_{0}\pi_{0}\ket{\alpha} + \bra{\alpha}\phi_{0}\pi_{0}\ket{\tilde{\alpha}} \right) \\
                &\qquad= \lim_{\alpha_0\rightarrow 0}\lim_{\tilde{\alpha}_0\rightarrow\alpha}\frac{1}{2\bra{\tilde{\alpha}_0}\ket{\alpha_0}} \left(\int d\kappa_0~\tilde{\alpha}_0\kappa_0\bra{\tilde{\alpha}_0}\ket{\kappa_0}\bra{\kappa_0}\ket{\alpha_0} + \int d\kappa_0~\alpha_0\kappa_0\bra{\alpha_0}\ket{\kappa_0}\bra{\kappa_0}\ket{\tilde{\alpha}_0} \right) \\
                &\qquad= \lim_{\alpha_0^+\rightarrow 0}\lim_{\alpha_0^-\rightarrow 0}\frac{1}{2\bra{\tilde{\alpha}_0}\ket{\alpha_0}} \left( \frac{-i}{4\pi}\int d\kappa_0 ~ e^{2\pi i \kappa_0 \alpha_0^-} + \frac{-i}{4\pi}\int d\kappa_0 ~ e^{2\pi i \kappa_0 \alpha_0^-}\right) \\
                &\qquad= \frac{-i}{4\pi}. \\
        \end{aligned}
    \end{equation}
    For more general modes, we have
    \begin{equation}
        \expval{\phi_{m}\pi_{n}} = \frac{-i}{4\pi}\delta_{m+n}, \qquad \expval{\pi_{m}\phi_{n}} = \frac{i}{4\pi}\delta_{m+n},
    \end{equation}
    which is consistent with the commutation relation $\expval{\phi_m\pi_n} - \expval{\pi_n\phi_m} = \expval{[\phi_m,\pi_n]} = \frac{-i}{2\pi}\delta_{m+n}$. The VEVs of two $\pi_n$ modes is divergent. Considering $\expval{\pi_0\pi_0}$ for example, we choose box normalization to regularize the divergence and find
    \begin{equation}
        \begin{aligned}
            &\expval{\pi_0\pi_0} 
                = \lim_{\alpha\rightarrow 0}\lim_{\tilde{\alpha}\rightarrow\alpha}\frac{1}{2\bra{\tilde{\alpha}}\ket{\alpha}} \left(\bra{\tilde{\alpha}}\pi_0\pi_0\ket{\alpha} + \bra{\alpha}\pi_0\pi_0\ket{\tilde{\alpha}} \right) \\
                & \qquad = \lim_{\alpha_0^+\rightarrow 0}\lim_{\alpha_0^-\rightarrow 0} \frac{\int d\kappa_0 ~ \kappa_0^2 \exp{2\pi i\kappa_0\alpha_0^-} + \kappa_0^2 \exp{-2\pi i\kappa_0\alpha_0^-}}{2\int d\kappa_0\exp{2\pi i\kappa_0\alpha_0^-}} \\
                & \xlongequal{\text{box normalization}} \lim_{\alpha_0^-\rightarrow 0} \lim_{K\rightarrow \infty} \frac{\int_{-K}^K d\kappa_0 ~\kappa_0\kappa_0 \exp{2\pi i\kappa_0\alpha_0^-} + \kappa_0\kappa_0 \exp{-2\pi i\kappa_0\alpha_0^-}}{2\int_{-K}^K d\kappa_0\exp{2\pi i\kappa_0\alpha_0^-}} \\
                & \xlongequal{\text{taking } \alpha_0^-\to 0 \text{ fist}} \lim_{K\rightarrow \infty} \frac{1}{3}K^2 \sim \frac{1}{\epsilon^2}.
        \end{aligned}
    \end{equation}
    Here we see the divergence of $\expval{\pi_0\pi_0} $ is $\epsilon^{-2}$. This divergence is intrinsic. To be more explicit, the uncertainty principle for a unitary theory requires
    \begin{equation}
        \begin{aligned}
            \expval{\phi_0^2}\expval{\pi_0^2} & \ge |\expval{\phi_0\pi_0}|^2 = \expval{\phi_0\pi_0}\expval{\pi_0\phi_0} = \frac{1}{4}((\expval{\phi_0\pi_0}+\expval{\pi_0\phi_0})^2-(\expval{\phi_0\pi_0}-\expval{\pi_0\phi_0})^2) \\
                & = (\Re{\expval{\phi_0\pi_0}})^2+\frac{1}{16\pi^2}\ge \frac{1}{8\pi} \\
        \end{aligned}
    \end{equation} 
    Thus as we taking $\expval{\phi_0\phi_0}\to 0$, $\expval{\pi_0\pi_0}\to \infty$ is divergent. This holds true for the modes with nonzero $n$ as well 
    \begin{equation}
        \expval{\pi_m\pi_n}\sim\frac{1}{\epsilon^2}\delta_{m+n}.
    \end{equation}
    However the coefficients in divergence depend on the normalization scheme, which lead to an arbitrary function in the correlator $\expval{\pi\pi}$.With the above results, we can read the 2-point correlators of the fields:
    \begin{equation}
        \begin{aligned}
            &\expval{\phi(\tau_1,\sigma_1)\phi(\tau_2,\sigma_2)} = 0, \\
            &\expval{\phi(\tau_1,\sigma_1)\pi(\tau_2,\sigma_2)} = -\frac{i}{2} \delta(\sigma_{12}),\\
            &\expval{\pi(\tau_1,\sigma_1)\phi(\tau_2,\sigma_2)} = \frac{i}{2} \delta(\sigma_{12}),\\
            &\expval{\pi(\tau_1,\sigma_1)\pi(\tau_2,\sigma_2)} = \frac{1}{\epsilon^2}f(\sigma_{12}) + \frac{i\tau_{12}}{2} \partial^2\delta(\sigma_{12}), \\
        \end{aligned}
    \end{equation}
    where $f$ is an arbitrary function. Here we have used $\tau_{12} = \tau_1-\tau_2$ and $\sigma_{12} = \sigma_1-\sigma_2$ for simplicity. By the identity that $\expval{\pi(\tau_1,\sigma_1)\pi(\tau_2,\sigma_2)}^* = \expval{\pi(\tau_2,\sigma_2)\pi(\tau_1,\sigma_1)}$, we see that $f$ satisfies $f^*(\sigma)=f(-\sigma)$. Thus the time-ordered propagators are
    \begin{equation}\label{eq:FeynmanPropagatorOf2DMagneticScalar}
        \begin{aligned}
            &\expval{\mathcal{T}\phi(\tau_1,\sigma_1)\phi(\tau_2,\sigma_2)} = 0, \\
            &\expval{\mathcal{T}\phi(\tau_1,\sigma_1)\pi(\tau_2,\sigma_2)} = -\frac{i}{2} \text{sign}(\tau_{12}) \delta(\sigma_{12}), \quad
            \expval{\mathcal{T}\pi(\tau_1,\sigma_1)\phi(\tau_2,\sigma_2)} = \frac{i}{2} \text{sign}(\tau_{12}) \delta(\sigma_{12}), \\
            &\expval{\mathcal{T}\pi(\tau_1,\sigma_1)\pi(\tau_2,\sigma_2)} = \frac{1}{\epsilon^2}(\theta(\tau_{12})f(\sigma_{12})+\theta(-\tau_{12})f(-\sigma_{12})) + \frac{i}{2} \abs{\tau_{12}}\partial^2\delta(\sigma_{12}),\\
        \end{aligned}
    \end{equation}
    where $\theta$ is the Heaviside step function. \par

\subsubsection{Path integral quantization}\label{subsubsec:2DPathIntegral}

    The other way to quantize the theory is by using the path integral. Now we need to first symmetrize the action \eqref{eq:Action2DMagneticScalar} by adding some total derivative terms,
    \begin{equation}
        \begin{aligned}
            S & = -\frac{1}{2} \int d^2\sigma ~ 2\pi \partial_\tau \phi + \partial_\sigma \phi \partial_\sigma \phi +\partial_\tau(-\pi\phi) + \partial_\sigma(-\phi\partial_\sigma\phi) \\
                & = \int d^2\sigma ~ \frac{1}{2} \phi \partial_\tau \pi -\frac{1}{2} \pi \partial_\tau \phi +\frac{1}{2} \phi \partial^2_\sigma \phi\\
                & = \int d^2\sigma ~ (\phi ~ \pi)\begin{pmatrix}\frac{1}{2}\partial_\sigma^2 & \frac{1}{2}\partial_\tau \\  -\frac{1}{2}\partial_\tau & 0\end{pmatrix}\begin{pmatrix}\phi \\ \pi \end{pmatrix}  
                = \int d^2\sigma ~ \mathbf{\Phi}^\dagger\hat{D}\mathbf{\Phi},\\
        \end{aligned}
    \end{equation}
    where 
    \begin{equation}
     \hat{D} = \begin{pmatrix}\frac{1}{2}\partial_\sigma^2 & \frac{1}{2}\partial_\tau \\  -\frac{1}{2}\partial_\tau & 0\end{pmatrix},\hspace{3ex}  
     \mathbf{\Phi} = \begin{pmatrix}\phi \\ \pi \end{pmatrix}.
    \end{equation}
    Thus the generating functional reads
    \begin{equation}
        \begin{aligned}
            Z[\mathbf{J}] & = \int \mathcal{D}\phi\mathcal{D}\pi ~ \exp{iS + i\int d^2\sigma ~ J_\phi\phi + J_\pi\pi } \\
                & = \int \mathcal{D}\phi\mathcal{D}\pi ~ \exp{i \int d^2\sigma ~ \mathbf{\Phi}^\dagger\hat{D}\mathbf{\Phi} + \frac{1}{2} \mathbf{J}^\dagger\mathbf{\Phi} + \frac{1}{2} \mathbf{\Phi}^\dagger \mathbf{J} } \\
                & = N \exp{i \int d^2\sigma_1 d^2\sigma_2 -\frac{1}{4} \mathbf{J}^\dagger(\sigma_1)\hat{D}^{-1}(\sigma_1-\sigma_2) \mathbf{J}(\sigma_2)}, \\
        \end{aligned}
    \end{equation}
    where $N$ is numerical coefficient given by the Gaussian integral, and $G(\sigma_1-\sigma_2) = \hat{D}^{-1}(\sigma_1-\sigma_2)$ is the Green's function of the operator $\hat{D}$. Using the Fourier transformation, we get 
    \begin{equation}\label{eq:GreenFunctionOf2DMagneticScalar}
        \begin{aligned}
            G(\sigma) &= \sum_{k\in\mathbb{Z}} \int \frac{d\omega}{(2\pi)^2}\begin{pmatrix} -\frac{k^2}{2} & -\frac{i\omega}{2}  \\ \frac{i \omega }{2} & 0 \\ \end{pmatrix}^{-1} e^{-i\omega\tau - i k\sigma}
                = \sum_{k\in\mathbb{Z}} \int \frac{d\omega}{(2\pi)^2}\begin{pmatrix} 0 & -\frac{2 i}{\omega } \\ \frac{2 i}{\omega } & \frac{2 k^2}{\omega ^2} \\\end{pmatrix}  e^{-i\omega\tau - i k\sigma} \\
                &= \begin{pmatrix} 0 & -\text{sign}(\tau)\delta(\sigma) \\ \text{sign}(\tau)\delta(\sigma) & A(\sigma) + \abs{\tau}\partial^2_\sigma\delta(\sigma) \\ \end{pmatrix}.
        \end{aligned}
    \end{equation}\par
    
\begin{figure}[ht]
        \centering
        \includegraphics[width=15cm,align=c]{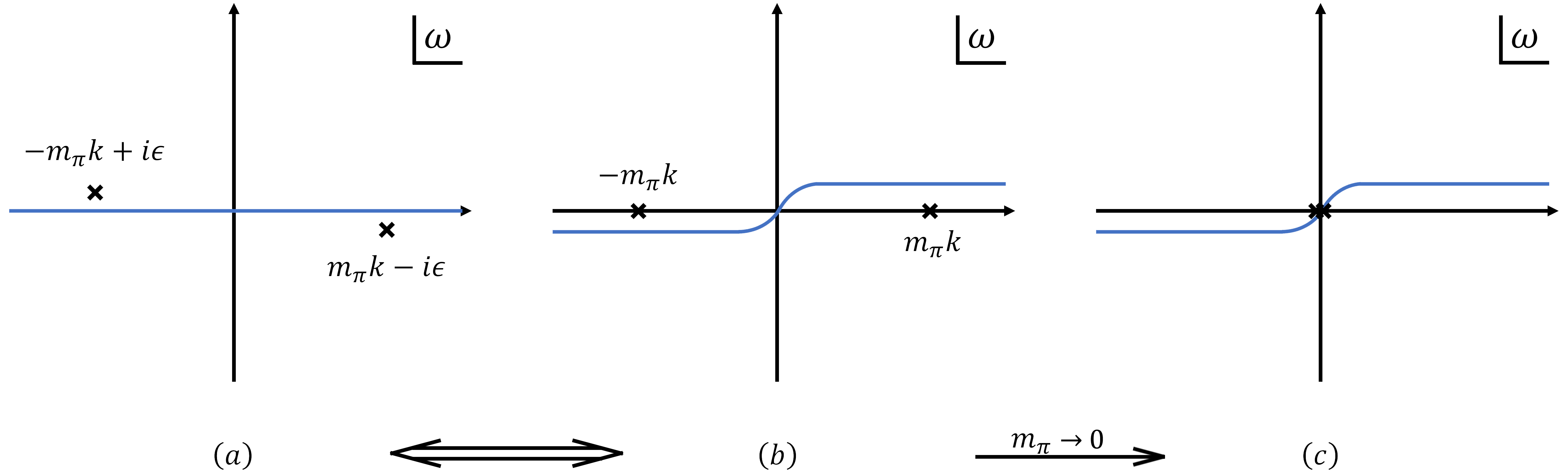}
        \caption{\centering In the complex $\omega$ plane, the contour of the integral in equation \eqref{eq:GreenFunctionOfPiPi2pt} is the blue curve in (a). This integral is equivalent to the integral of Feynman propagator along the contour shown in (b). In taking the $m_\pi \to 0$ limit, the two poles merge into one, and the contour shown in (c) is taken as running through the pole.} 
        \label{fig:FyenmanIntigralCurve}
    \end{figure}

\begin{figure}[ht]
        \centering
        \includegraphics[width=5cm,align=c]{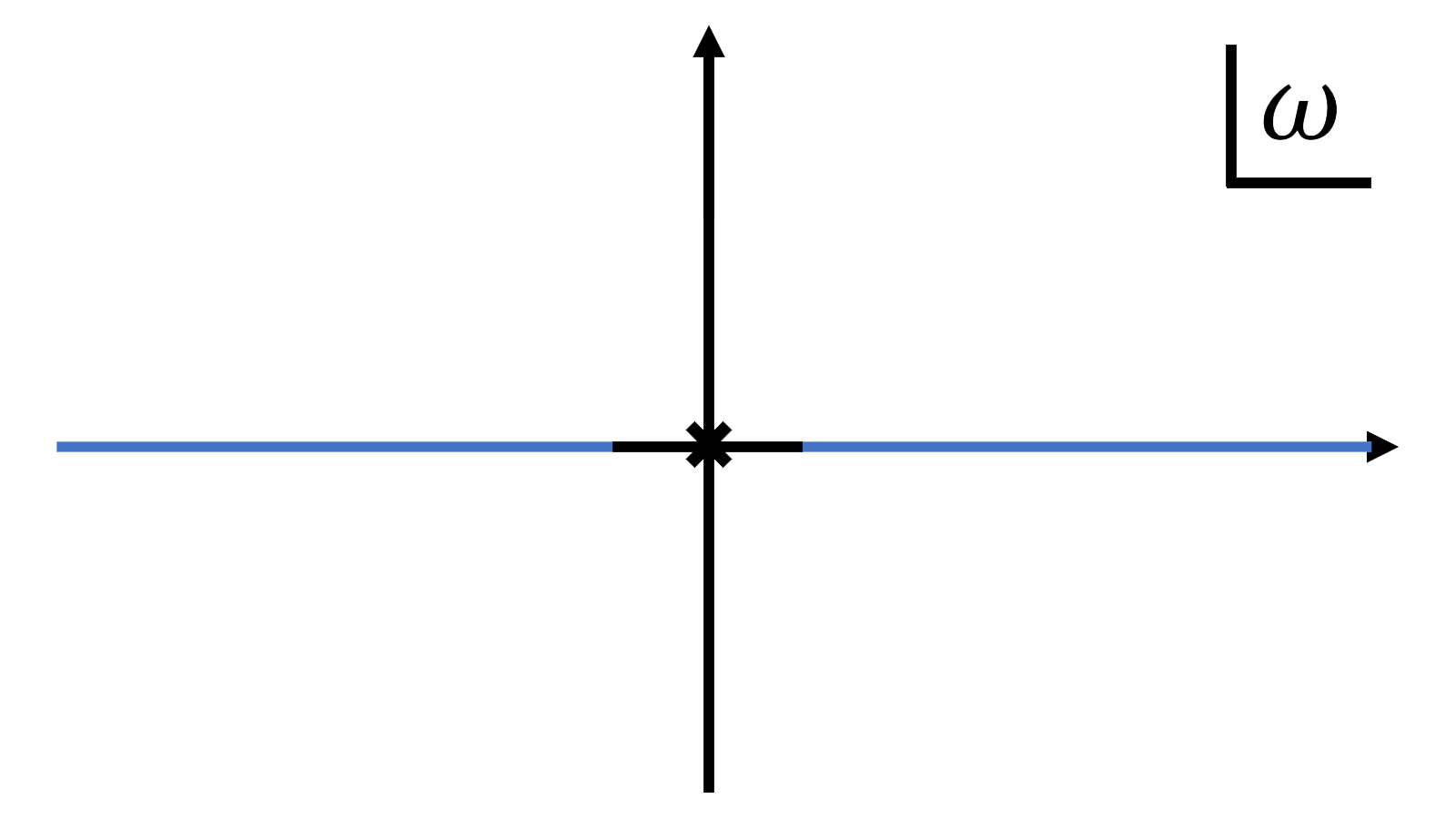}
        \caption{\centering The contour of the principal-valued integral in the complex $\omega$ plane. }
        \label{fig:principalIntigralCurve}
    \end{figure}

    The function $A(\sigma)$ needs some clarifications.  It stems from the integral contour for the Feynman propagator, i.e. time-ordered propagator. To see this term, we should add a mass term of $\pi$ field, $- \frac{1}{2} m_\pi^2\pi\pi$, to the Lagrangian\footnote{Although this mass term $\frac{1}{2}m_\pi^2\pi\pi$ breaks the Carrollian symmetry, it is very helpful in calculation. The mass $m_\pi$ will be eventually taken to be $0$ to recover the symmetry. The mass term of $\phi$ change $k^2$ term to $k^2 + m_\phi^2$, which gives no help in calculation. }. This mass term modifies the derivative operator. In the momentum space, the modified derivative operator and Green's function reads
    \begin{equation}
        \begin{aligned}
            &\tilde{\hat{D}} = \sum_{k\in\mathbb{Z}} \int \frac{d\omega}{(2\pi)^2}\begin{pmatrix} -\frac{k^2}{2} & -\frac{i\omega}{2}  \\ \frac{i \omega }{2} & -\frac{m_\pi^2}{2} \\ \end{pmatrix} e^{-i\omega\tau - i k\sigma}, \\
            &\tilde{G}(\sigma) = \sum_{k\in\mathbb{Z}} \int \frac{d\omega}{(2\pi)^2}\begin{pmatrix} \frac{2 m_\pi^2}{\omega ^2 - m_\pi^2 k^2} & -\frac{2 i\omega }{\omega ^2 - m_\pi^2 k^2} \\ \frac{2 i\omega }{\omega ^2 - m_\pi^2 k^2} & \frac{2 k^2}{\omega ^2 - m_\pi^2 k^2} \\\end{pmatrix}  e^{-i\omega\tau - i k\sigma}. \\
        \end{aligned}
    \end{equation}
    Doing the integral along the contour  as shown in Figure \ref{fig:FyenmanIntigralCurve}, we have
    \begin{equation}\label{eq:GreenFunctionOfPiPi2pt}
        \begin{aligned}
            \tilde{G}_{\pi\pi}(\sigma) &= \sum_{k\in\mathbb{Z}} \int_\mathcal{C} d\omega \frac{2 k^2}{\omega ^2 - m_\pi^2 k^2}  e^{-i\omega\tau - i k\sigma} \\
            & = \sum_{k\in\mathbb{Z}} \int d\omega \frac{2 k^2}{(\omega - (m_\pi k - i\epsilon))(\omega + (m_\pi k + i\epsilon))}  e^{-i\omega\tau - i k\sigma} \\
            & = \sum_{k\in\mathbb{Z}} \frac{-i k}{m_\pi} e^{-im_\pi k\abs{\tau}}  e^{- i k\sigma} \\
            & = \frac{1}{m_\pi} \partial\delta(\sigma + m_\pi \abs{\tau}) 
             \xlongequal{m_\pi=\epsilon^2\to 0} \frac{1}{\epsilon^2}f(\sigma) = A(\sigma). \\
        \end{aligned}
    \end{equation}
    In the last step,  taking $m_\pi\to 0$ gives a power-law divergence times a function of $\sigma$. In fact, one can choose different ways to regularise this divergent integral. For example, $-\frac{1}{2}m_\pi \pi\partial_\sigma^2\pi$ results in $\tilde{G}_{\pi\pi}(\sigma) \sim \frac{1}{m_\pi} \delta(\sigma)$. Nevertheless, the Feynman propagator of $\pi(\sigma_1)\pi(\sigma_2)$ leads to a power-law-divergent function of $\sigma_1-\sigma_2$. Using the Hadamard regularization to subtract the divergence, we get the regular term 
    \begin{equation}
        \begin{aligned}
            G^{\text{reg}}_{\pi\pi}(\sigma) & = \sum_{k\in\mathbb{Z}} \int_{P.V.} d\omega ~ \frac{2 k^2}{\omega ^2}  e^{-i\omega\tau - i k\sigma} = \sum_{k\in\mathbb{Z}} 2 k^2\abs{\tau}  e^{- i k\sigma} = \abs{\tau}\partial^2\delta(\sigma), \\
        \end{aligned}
    \end{equation}
    which corresponds to the principal value. The contour of the principal-valued integral is shown in Figure \ref{fig:principalIntigralCurve}. Thus the the full integral  is given by
    \begin{equation}
        G_{\pi\pi}(\sigma)  = A(\sigma) + \abs{\tau}\partial^2\delta(\sigma) .
    \end{equation}
    For other components of $G$, we find that there is no pole in $m_\pi$ appearing in the correlator, and after taking the $m_\pi\to 0$ limit we read \eqref{eq:GreenFunctionOf2DMagneticScalar}.  \par

    Now we can finally calculate the correlators. We have
    \begin{equation}
        \expval{\mathbf{\Phi}_i(\sigma_1)\mathbf{\Phi}_j(\sigma_2)} = \left.\frac{1}{Z[\mathbf{J}]}\frac{\delta}{i\delta \mathbf{J}_i(\sigma_1)}\frac{\delta}{i\delta \mathbf{J}_j(\sigma_2)}Z[\mathbf{J}]\right|_{\mathbf{J} = 0}.
    \end{equation}
    Straightforward calculations shows that they  match exactly with the time-ordered correlation functions in the canonical quantization \eqref{eq:FeynmanPropagatorOf2DMagneticScalar}:
    \begin{equation}
        \begin{aligned}
            \expval{\mathbf{\Phi}_i(\sigma_1)\mathbf{\Phi}_j(\sigma_2)} &= \frac{i}{4}(G_{ij}(\sigma_{12}) + G_{ji}(\sigma_{21}))  \\
                & = \begin{pmatrix} 0 & -\frac{i}{2}\text{sign}(\tau_{12})\delta(\sigma_{12}) \\ \frac{i}{2}\text{sign}(\tau_{12})\delta(\sigma_{12}) & A(\sigma_{12}) + \frac{i}{2}\abs{\tau_{12}}\partial^2_\sigma\delta(\sigma_{12}) \\ \end{pmatrix} .\\
        \end{aligned}
    \end{equation}\par

\subsubsection{Hilbert space}\label{subsec:HilbertSpace}

    As defined in section \ref{subsubsec:CanonicalQuantizationOf2DMagneticScalar}, the Hilbert space is made of  infinite copies of $L^2(\mathbb{R})$ with basis $\ket{\alpha}$. Similar to what we learnt in quantum mechanics, $\ket{\alpha}$ is not in the Hilbert space, since all of them are not normalizable. But all the states in the Hilbert space can be decomposed into the basis states $\ket{\alpha}$, thus it is enough to consider the basis states. The basis states are related to the vacuum by
    \begin{equation}
        \ket{\alpha} = N\exp{-2\pi i\sum_{n}\alpha_{n}\pi_{-n}}\ket{0} = V(\alpha)\ket{0},
    \end{equation}
    where $N$ is the normalization factor. When all the components of $\alpha$ are equal $\alpha_n = \alpha_0$, the operator $V(\alpha_0)$ is a vertex operator which can be realized by inserting the operator at the origin $(\tau = 0,\sigma = 0)$.
    \begin{equation}
        V(\alpha_0) = N\exp{-2\pi i \alpha_0 ~\pi(\tau=0,\sigma=0)}= N\exp{-2\pi i\sum_{n}\alpha_{0}\pi_{n}}.
    \end{equation}
    For generic constant $\alpha$, the operator $V(\alpha)$ is generated by line operators at $\tau = 0$:
    \begin{equation}
        V(\alpha) = N\exp{- i \int d\sigma ~ \alpha_{n} \pi(\tau=0,\sigma)e^{-in\sigma}}= N\exp{-2\pi i\sum_{n}\alpha_{n}\pi_{-n}}.
    \end{equation}
    The vertex operator $V(\alpha_0)$ is a special case of $V(\alpha)$ which degenerate to the origin. Thus the basis states of the Hilbert space correspond to line operators $V(\alpha)$ rather than local vertex operators.  \par

    Furthermore, it seems that the discussions of the state-operator correspondence in CFT can not be directly extended to the Carrollian case. In the conformal case, there is cylinder-to-plane map,  which maps the past-infinity time slice on the cylinder to the origin on the plane. This means that a state in the Hilbert space of the cylinder is equivalent to  a local operator inserting at the origin on the plane.  On the other hand, as discussed in section \ref{subsec:BMS3Symmetry} for the Carrollian case, there is no transformation that maps the past-infinity equal-time slice $\tau=-\infty$ to a point. Thus the state-operator correspondence can not be expected in Carrollian conformal field theory.  This discussion is also valid for higher dimensional cases. \par

\subsection{Non-unitary canonical quantization}\label{subsec:2DHighestWeightQuantization}

    In the process of quantization, it is possible to select different vacua by relaxing some requirements. In this section, we consider the highest-weight vacuum, and the price to pay is the loss of the unitarity. As introduced in \cite{Bagchi:2020fpr}, there are multiple vacuum conditions. The Hilbert space defined in section \ref{subsubsec:CanonicalQuantizationOf2DMagneticScalar} realize the induced vacuum in the literature, while the vacuum condition in this section realize the flipped vacuum. \par

    To be precise, the condition on the highest-weight vacuum is 
    \begin{equation}
        \begin{aligned}
            L_{n\ge 1}\ket{\text{vac}}_h = M_{n\ge 1}\ket{\text{vac}}_h = 0,
        \end{aligned}
    \end{equation}
    while $L_{n\le -1}\ket{\text{vac}}_h\neq 0$ and $M_{n\le -1}\ket{\text{vac}}_h\neq 0$ give descendent states of the vacuum. The subscript $h$ stands for the highest weight. \par

\subsubsection{Non-unitarity of the highest-weight states}

    Consider a highest-weight state $\ket{h}$ satisfying 
    \begin{equation}\label{eq:2DHighestWeightStateCondition}
        \begin{aligned}
            L_{n\ge 1}\ket{h} = M_{n\ge 1}\ket{h} = 0.
        \end{aligned}
    \end{equation}
    In this subsection we prove that the nontrivial highest-weight states are non-unitary. Here ``nontrivial" means that $L_{n\le -1}\ket{h}\neq 0$ and $M_{n\le -1}\ket{h}\neq 0$, which generate descendent states. \par

    In the theory at hand, only $\phi_n$ and $\pi_n$ are the fundamental modes acting on states. The commutation relations and the conjugation relations of $\phi_n$ and $\pi_n$ modes tell that the modes can be decomposed into  infinite sets of operators $\{\phi_0, \pi_0\}$ and $\{\phi_n, \pi_n, \phi_{-n}, \pi_{-n}\}$ for $n\ge 1$. Thus the Hilbert space $\mathcal{H}$ can be taken as direct product of sub-Hilbert spaces $\mathcal{H} = \mathcal{H}_0 \otimes \mathcal{H}_1\otimes \mathcal{H}_2\otimes\cdots$ with $\{\phi_0, \pi_0\}$ acting on $\mathcal{H}_0$ and $\{\phi_n, \pi_n, \phi_{-n}, \pi_{-n}\}$ acting on $\mathcal{H}_n$. A state in $\mathcal{H}$ can be decomposed as $\ket{\psi} = \ket{\psi_0}\otimes\ket{\psi_1}\otimes\ket{\psi_2}\otimes\cdots$. The condition on the highest-weight state $\ket{h}$ \eqref{eq:2DHighestWeightStateCondition} can be realized in  two different ways:
    \begin{equation}
        \begin{aligned}
            \begin{aligned}
                &(i)    && \phi_{n\neq 0}\ket{h} = 0, \\
                &(ii)   && \phi_{n\ge 1}\ket{h} = \pi_{n\ge 1}\ket{h} = 0. \\
            \end{aligned}
        \end{aligned}
    \end{equation}\par

    The condition $(i)$ leads to trivial highest-weight state
    \begin{equation}
        \begin{aligned}
            \ket{h}=\ket{\psi_0}\otimes(\ket{\alpha^c_1=0}\otimes\ket{\alpha^s_1=0})\otimes\cdots \qquad
             L_{n}\ket{h} = M_{n}\ket{h} = 0 \quad n\in \mathbb{Z},
        \end{aligned}
    \end{equation}
    where $\ket{\psi_0}$ is an arbitrary state in $\mathcal{H}_0$, and $\ket*{\alpha^{c/s}_{n\ge 1}=0}$ are the eigenstates of $\phi^{c/s}_{n\ge 1}$ modes and $\ket{\psi_n} = \ket{\alpha^c_n=0}\otimes\ket{\alpha^s_n=0} \in\mathcal{H}_n $. The  state $\ket{h}$ is an excited state from the canonical vacuum which transforms trivially under the BMS$_3$ symmetry. It is a trivial highest-weight state because $L_{-n}$ and $M_{-n}$ would not generate new descendent states. \par
    
    On the other hand, the condition $(ii)$ leads to non-unitarity of the Hilbert space. Assuming the Hilbert space is unitary, for given $n\ge 1$ and arbitrary state $\ket{\psi}\in\mathcal{H}$, we have
    \begin{equation}
        \begin{aligned}
            &\bra{h}\phi_n\phi_{-n}\ket{h}\bra{\psi}\ket{\psi} = ||\phi_{-n}\ket{h}||^2||\ket{\psi}||^2 \ge |(\ket{\psi}, \phi_{-n}\ket{h})|^2 ,
        \end{aligned}
    \end{equation} 
    where we have used the Cauchy inequality. Since $[\phi_m, \phi_n] = 0$, we know that $\bra{h}\phi_n\phi_{-n}\ket{h} = \bra{h}\phi_{-n}\phi_n\ket{h} = 0$. Thus
    \begin{equation}
        |(\ket{\psi}, \phi_{-n}\ket{h})|^2 = 0 ,
    \end{equation}
    which means $\phi_{-n}\ket{h}$ is a null state. For the same reason  $\pi_{-n}\ket{h}$ is a null state as well. Moreover, the commutation relations lead to
    \begin{equation}
        \begin{aligned}
            &\frac{-i}{2\pi}\ket{h} =[\phi_{n}, \pi_{-n}]\ket{h} = \phi_{n}\pi_{-n}\ket{h} - \pi_{-n}\phi_{n}\ket{h} = |null\rangle, \\
            &\frac{i}{2\pi}\ket{h} =[\pi_{n}, \phi_{-n}]\ket{h} = \pi_{n}\phi_{-n}\ket{h} - \phi_{-n}\pi_{n}\ket{h} = |null\rangle, \\
        \end{aligned}
    \end{equation}
    which is a contradiction. Thus in conclusion, the non-trivial highest-weight states break the unitarity of the Hilbert space. Nevertheless, we can choose the vacuum $\ket{\text{vac}}_h$ to be  a nontrivial highest-weight state and give up the unitarity. 
    \par

\subsubsection{Canonical quantization with the highest-weight vacuum}

    We define the highest-weight vacuum in a non-unitary Hilbert space. The vacuum $\ket{\text{vac}}_h$ and $_h\bra{\text{vac}}$ are defined as 
    \begin{equation}\label{eq:2DHighestWeightVacuumCondition}
        \begin{aligned}
            &\phi_{n>0}\ket{\text{vac}}_h = \pi_{n\ge 0}\ket{\text{vac}}_h = 0, 
            &&_h\bra{\text{vac}}\phi_{n\le 0} = _h\bra{\text{vac}}\pi_{n<0} = 0. \\
        \end{aligned}
    \end{equation}
    Notice that $\pi_0$ only annihilates the in-vacuum $\ket{\text{vac}}_h$ but not the out-vacuum $_h\bra{\text{vac}}$. The excited states are generated by $\phi_{n\le0}$ and $\pi_{n<0}$ modes acting on the vacuum. \par

    For this vacuum, the normal ordering for non-zero modes is to put the positive modes to the right, the negative modes to the left, and $\phi_0$ mode to the left of $\pi_0$ mode. After the normal ordering,  $L_0$ takes the following form
    \begin{equation}
        \begin{aligned}
            L_0 = \sum_{a>0} 2\pi i a(\phi_{-a}\pi_{a} - \pi_{-a}\phi_a).
        \end{aligned}
    \end{equation}
    Other generators have no ordering ambiguity. Thus the vacuum is invariant under
    \begin{equation}
        \begin{aligned}
            L_{n\ge-1}\ket{\text{vac}}_h &= M_{n\ge-1}\ket{\text{vac}}_h = 0,\\
            _h\bra{\text{vac}}L_{n\le-1} &= _h\bra{\text{vac}}M_{n\le-1} = 0. \\
        \end{aligned}
    \end{equation} 
 In this vacuum, there are non-trivial anomalous terms in the commutation relations among the symmetry generators, 
    \begin{equation}\relax 
        \begin{aligned}
            &[L_m, L_n] = (m-n)L_{m+n} + a_L(m)\delta_{m+n}, \\& [L_m, M_n] = (m-n)M_{m+n} + a_M(m)\delta_{m+n},
        \end{aligned}
    \end{equation}
    where $a_L$ and $a_M$ are the anomalous terms.
    Using the Jacobi identity, we get 
    \begin{equation}
        \begin{aligned}
            &a_L(n) = c^L_1 n^3 + c^L_2 n, && a_M(n) = c^M_1 n^2 + c^M_2.
        \end{aligned}
    \end{equation}
    Further considering VEVs of the commutators of the generators, we get
    \begin{equation}
        \begin{aligned}
            &a_L(n) = \frac{1}{6} (n^3 - n), && a_M(n) = 0.
        \end{aligned}
    \end{equation} 
    The anomaly $a_L$ has a similar form to the one in $2$D CFT. \par

    Using the definition of the vacuum \eqref{eq:2DHighestWeightVacuumCondition}, we can easily read the 2-point correlators:
    \begin{equation}
        \begin{aligned}
            &\expval{\phi(\tau_1,\sigma_1)\phi(\tau_2,\sigma_2)}  = 0, \\
            &\expval{\phi(\tau_1,\sigma_1)\pi(\tau_2,\sigma_2)} = \frac{-i}{2\pi} \frac{1}{e^{i\sigma_{12}} - 1}, \\
            &\expval{\pi(\tau_1,\sigma_1)\phi(\tau_2,\sigma_2)} =  \frac{-i}{2\pi} \frac{1}{e^{-i\sigma_{12}} - 1},  \\
            &\expval{\pi(\tau_1,\sigma_1)\pi(\tau_2,\sigma_2)}  = \frac{-i\tau_{12}}{2\pi} \frac{e^{i\sigma_{12}}(e^{i\sigma_{12}} + 1)}{(e^{i\sigma_{12}} - 1)^2}. \\
        \end{aligned} 
    \end{equation}
    These propagators can not be read by using the path integral. These correlation functions are of power-law forms in plane, which match the structures of those by taking the $c\to0$ limit of the correlators in a CFT. To see this, we can map the theory to a plane by a BMS$_3$ transformation, 
    \begin{equation}
        \begin{aligned}
            &t= i\tau e^{i\sigma}, \qquad x= e^{i\sigma}, \\
            &\tilde{\phi}(t,x) = \phi(\tau,\sigma),\\
            &\Tilde{\pi}(t,x) = -ie^{-i\sigma}\left(\pi(\tau, \sigma) + i\tau \partial_\sigma\phi(\tau, \sigma) +\frac{\tau^2}{2}\partial_\tau\phi(\tau, \sigma) \right).\\
        \end{aligned}
    \end{equation}
    The correlation functions on the plane are of power-law forms in the plane coordinates. For example, 
    \begin{equation}
        \begin{aligned}
            \expval*{\tilde{\phi}(t_1,x_1)\Tilde{\pi}(t_2,x_2)} &= -\frac{1}{2\pi}\frac{1}{x_{12}}, &&
            \expval*{\tilde{\pi}(t_1,x_1)\Tilde{\pi}(t_2,x_2)} &= \frac{1}{\pi}\frac{t_{12}}{(x_{12})^3}. \\
        \end{aligned}
    \end{equation}

\section{Magnetic scalar in \texorpdfstring{$\mathbb{R}\times\mathbb{R}^2$}{R×R2}}\label{sec:3DMagneticScalar}

    Using almost the same method with the one in section \ref{sec:2DMagneticScalar}, we can discuss the $d=3$ magnetic scalar theory with BMS$_4$ symmetry. The BMS$_4$ symmetry in principle can be realized on any Carrollian manifold $\mathbb{R}\times\mathcal{R}^2$ with $\mathcal{R}^2$ being a generic Riemann manifold. But in practice, it is hard to consider the BMS$_4$ transformations on the simplest $\mathbb{R}\times S^2$, the $3$D analog of cylinder. The reason is that in the $S^2$ part the orthogonal basis is the spherical oscillators. This basis is covariant under the $\mathfrak{so}(3)=\{L_{0,\pm1}, \bar{L}_{0,\pm1}\}$ part of the BMS$_4$ algebra, but it is not closed under the actions of other transformations in BMS$_4$. Thus in this section, we discuss the magnetic scalar theory on $\mathbb{R}\times\mathbb{R}^2$. \par

\subsection{BMS\texorpdfstring{$_4$}{4} symmetry on \texorpdfstring{$\mathbb{R}\times\mathbb{R}^2$}{R×R2} }

    Consider the $3$-dimensional Carrollian manifold consisting of $\mathbb{R}\times\mathbb{R}^2$ with coordinates $x^\mu = (x^0=t, x^1, x^2)$, a degenerated metric, and a time-like vector taken to be
    \begin{equation}\label{eq:3DMetricAndTimeLikeVector}
        g=\begin{pmatrix}0&0&0\\ 0& 1& 0\\0&0&1\end{pmatrix},\qquad \zeta = (1,0,0).
    \end{equation}
    The BMS$_4$ symmetry transformations keep $g$ and $\zeta$ invariant. Thus the Carrollian conformal Killing equations on the vector $\xi^\mu=(\xi^0,\xi^1,\xi^2)$ are
    \begin{equation}
        \begin{aligned}
            &\partial_0\xi^0 = \partial_1\xi^1 = \partial_2\xi^2, \quad \partial_0\xi^1 = \partial_0\xi^2 = 0, \quad \partial_2\xi^1 = -\partial_1\xi^2.
        \end{aligned}
    \end{equation}
    The solutions of $\xi^1$ and $\xi^2$ are conjugate harmonic functions on $(x^1, x^2)\in \mathbb{R}^2$. Therefore, it is better to solve the equations in terms of the complexified coordinates $(t, z, \bar{z})$ with $z = x^1+ix^2$ and $\bar{z} = x^1-ix^2$. The Carrollian conformal Killing equations are now
    \begin{equation}
        \begin{aligned}
            &\partial_t\xi^t = \frac{1}{2}\partial\xi + \frac{1}{2}\partial\bar{\xi}, 
                \quad \partial_t\xi =\partial_t\bar{\xi} = 0, 
                \quad \partial\bar{\xi} = \bar{\partial}\xi = 0. \\
        \end{aligned}
    \end{equation}
    Here $\partial=\partial_z$, $\bar{\partial}=\partial_{\bar{z}}$, and the components of the conformal Killing vector are $\xi = \xi^1 + i\xi^2$, $\bar{\xi} = \xi^1 - i\xi^2$. The solution is given simply by
    \begin{equation}
        \begin{aligned}
                & \xi^t = \frac{1}{2}(\partial f(z) + \bar{\partial}\bar{f}(\bar{z}) ) t + g(z,\bar{z}),
                && \xi = f(z), 
                && \bar{\xi} = \bar{f}(\bar{z}), \\
        \end{aligned}
    \end{equation}
    where $\bar{f} (\bar{z})= (f(z))^*$ is the complex conjugation of $f$, and $g$ is an arbitrary real function. Thus the coordinates transform infinitesimally and finitely as 
    \begin{equation}\label{eq:BMS4Transformation}
        \begin{aligned}
            &\left\{\begin{aligned} 
                & \delta t = \xi^t = \frac{t}{2}(\partial f(z) + \bar{\partial}\bar{f}(\bar{z}) ) + g(z,\bar{z})\\ 
                & \delta z = \xi = f(z) \\ 
                & \delta \bar{z} = \bar{\xi} = \bar{f}(\bar{z}) \end{aligned}\right. ,
            & \left\{\begin{aligned} 
                & \Tilde{t} = (\partial F(z) \bar{\partial}\bar{F}(\bar{z}))^{\frac{1}{2}} t + G(z,\bar{z})\\ 
                & \Tilde{z} = F(z)\\ 
                & \Tilde{\bar{z}} = \bar{F}(\bar{z}) \end{aligned}\right. . \\
        \end{aligned}
    \end{equation}
    The BMS$_4$ symmetry is generated by three sets of infinite number of generators,
    \begin{equation}\label{eq:BMS4GeneratingVectors}
        l_n = -\frac{n+1}{2}t z^n\partial_t - z^{(n+1)}\partial_z, 
        \quad \bar{l}_{\bar{n}} = -\frac{\bar{n}+1}{2}t \bar{z}^{\bar{n}}\partial_t - \bar{z}^{(\bar{n}+1)}\partial_{\bar{z}}, 
        \quad m_{n,\bar{n}} =  -z^{n}\bar{z}^{\bar{n}}\partial_t,
    \end{equation}
    with the commutation relations
    \begin{equation}\label{eq:BMS4Algebra}
        \begin{aligned}
            &[l_m,l_n] = (m-n) l_{m+n}, \qquad [\bar{l}_{\bar{m}},\bar{l}_{\bar{n}}] = (\bar{m}-\bar{n}) \bar{l}_{m+n}, \qquad [l_{m},\bar{l}_{\bar{m}}] = 0, \\
            &[l_m,m_{n,\bar{n}}] = \left(\frac{m+1}{2} - n \right)m_{m+n,\bar{n}}, \qquad [\bar{l}_{\bar{m}},m_{n,\bar{n}}] = \left(\frac{\bar{m}+1}{2} - \bar{n} \right)m_{n,\bar{m}+\bar{n}} , \\
            &[m_{m,\bar{m}},m_{n,\bar{n}}] = 0 .
        \end{aligned}
    \end{equation}
    Since the BMS$_4$ generators transform the real coordinates $x^\mu$ to complex ones,  only the combinations of BMS$_4$ generators which generate real transformations correspond to the symmetry transformations of the Carrollian manifold. \par

\subsection{Classical symmetries of the \texorpdfstring{$d=3$}{d=3} magnetic Carrollian scalar theory}

    The BMS$_4$-invariant magnetic scalar theory is given by
    \begin{equation}\label{eq:Action3DMagneticScalar}
        S = - \frac{1}{2}\int d^3x ~ 2\pi \partial_t \phi + \partial_i \phi \partial_i \phi ,
    \end{equation}
    with $i=1,2$. The fundamental fields are $\pi$ and $\phi$, and the corresponding equations of motion can be read from the action \eqref{eq:Action3DMagneticScalar},
    \begin{equation}
        \begin{aligned}
            \pi: & &&\partial_t \phi = 0, \\
            \phi: & &&\partial_t \pi + \partial_i \partial_i \phi = 0. \\
        \end{aligned}
    \end{equation}
    The fields can be  expanded in the Fourier basis as
    \begin{equation}
        \begin{aligned}
            &\phi(x) = \int d^2\vec{k} ~ \phi(\vec{k}) e^{-i \vec{k}\cdot\vec{x}}, \\
            &\pi(x) = \int d^2k ~ (\pi(\vec{k}) + t\vec{k}^2\phi(\vec{k})) e^{-i \vec{k}\cdot\vec{x}}, \\
        \end{aligned}
    \end{equation} 
    where $\vec{x} = (x^1, x^2)$ and $\vec{k} = (k^1, k^2)$. \par
    
    Under the BMS$_4$ transformations \eqref{eq:BMS4Transformation}, the fields transform as 
    \begin{equation}
        \begin{aligned} 
            &\Tilde{\phi}(\Tilde{x}) = (\partial F \bar{\partial}\bar{F})^{-\frac{1}{4}}\phi(x), \\
            &\Tilde{\pi}(\Tilde{x}) = (\partial F \bar{\partial}\bar{F})^{-\frac{3}{4}} \left\{\pi(x) + \left(\frac{t\bar{\partial}^2\bar{F}}{2\bar{\partial}\bar{F}} + \frac{\bar{\partial} G}{(\partial F \bar{\partial}\bar{F})^{\frac{1}{2}}} \right) \partial\phi(x) + \left(\frac{t\partial^2 F}{2\partial F} + \frac{\partial G}{(\partial F \bar{\partial}\bar{F})^{\frac{1}{2}}} \right) \bar{\partial}\phi(x)\right. \\
                & \qquad\left. - \left(\frac{t\partial^2 F}{2\partial F} + \frac{\partial G}{(\partial F \bar{\partial}\bar{F})^{\frac{1}{2}}} \right) \left(\frac{t\bar{\partial}^2\bar{F}}{2\bar{\partial}\bar{F}} + \frac{\bar{\partial} G}{(\partial F \bar{\partial}\bar{F})^{\frac{1}{2}}} \right) \partial_t\phi(x) +\left(\frac{\partial\bar{\partial}G}{2\partial F \bar{\partial}\bar{F} }-\frac{t\partial^2 F\bar{\partial}^2\bar{F}}{8\partial F \bar{\partial}\bar{F}} \right) \phi(x) \right\}.  \\
        \end{aligned} \nonumber
    \end{equation}
    Here we used $\tilde{x} = (\Tilde{t}, \Tilde{x}^1, \Tilde{x}^2)$ as the transformed coordinates for simplicity. Infinitesimally, the fields transform as
    \begin{equation}\label{eq:3DInfinitesimalBMS4Transformation}
        \begin{aligned} 
            & \delta \phi(x) = \Tilde{\phi}(\tilde{x}) - \phi(x) = -\frac{1}{2}\partial_t\xi^0\phi , \\ 
            & \delta \pi(x) = \tilde{\pi}(\tilde{x}) - \pi(x) = -\frac{3}{2}\partial_t\xi^0 \pi + \frac{1}{2}\partial_i\xi^0 \partial_i\phi + \frac{1}{4}\partial_i \partial_i \xi^0 \phi . \\
        \end{aligned} 
    \end{equation}
    It can be checked that the theory \eqref{eq:Action3DMagneticScalar} is invariant under the BMS$_4$ symmetry up to total derivatives,
    \begin{equation}
        \begin{aligned}
            \tilde{S} &= S - \int d^3x ~ (\text{total derivatives}).
        \end{aligned}
    \end{equation}
    
    The stress tensor for the 3D magnetic scalar theory is given by
    \begin{equation}
        T^\mu_{~\nu} = \left(\begin{smallmatrix} \scriptstyle
            - \frac{1}{4}\partial_i\phi\partial_i\phi + \frac{1}{4}\phi\partial_i\partial_i\phi & \frac{3}{4}\pi\partial_1\phi - \frac{1}{4}\phi\partial_1\pi & \frac{3}{4}\pi\partial_2\phi - \frac{1}{4}\phi\partial_2\pi\\ 
            0 & \frac{1}{2} \partial_1\phi\partial_1\phi + \frac{1}{4} \partial_2\phi\partial_2\phi - \frac{1}{4} \phi\partial_1\partial_1\phi & \frac{3}{4}\partial_1\phi\partial_2\phi - \frac{1}{4}\phi\partial_1\partial_2\phi \\ 
            0 & \frac{3}{4}\partial_1\phi\partial_2\phi - \frac{1}{4}\phi\partial_1\partial_2\phi & \frac{1}{4} \partial_1\phi\partial_1\phi + \frac{1}{2} \partial_2\phi\partial_2\phi - \frac{1}{4} \phi\partial_2\partial_2\phi \end{smallmatrix}\right),
    \end{equation}
    satisfying the conserved current equation $\partial_\mu T^\mu_{~\nu} = 0$, the traceless condition $\Tr T^\mu_{~\nu} = 0$, and the Carrollian stress tensor structure $T^i_{~0} = 0$, $T^i_{~j} = T^j_{~i}$. Thus the charges $Q_\xi = -i\int d^2x ~ \xi^\mu T^0_{~\mu}$ generated by the Carrollian conformal Killing vectors $\xi^\mu$ are conserved,
    \begin{equation}
        \begin{aligned}
            \partial_0 Q_\xi & = -i \int d^2x ~ \partial_0\xi^\mu T^0_{~\mu} + \xi^\mu \partial_0 T^0_{~\mu} = -i \int d^2x ~ \partial_i(\xi^\mu T^i_{~\mu}) = 0.\\
        \end{aligned}
    \end{equation}
    Thus the theory is classically Carrollian conformal invariant. The BMS$_4$ charges are defined by
    \begin{equation}\label{eq:BMS4GeneratorsInStressTensor}
        \begin{aligned}
            &L_n = i\int d^2x ~ \frac{n+1}{2}t(x^1+ix^2)^n T^0_{~0} + \frac{1}{2}(x^1+ix^2)^{(n+1)} (T^0_{~1} - iT^0_{~2}), \\
            &\bar{L}_{\bar{n}} = i\int d^2x ~ \frac{\bar{n}+1}{2}t(x^1-ix^2)^{\bar{n}} T^0_{~0} + \frac{1}{2}(x^1+ix^2)^{(\bar{n}+1)} (T^0_{~1} + iT^0_{~2}), \\
            &M_{n,\bar{n}} = i \int d^2x ~ (x^1+ix^2)^n (x^1-ix^2)^{\bar{n}} T^0_{~0} . \\
        \end{aligned}
    \end{equation} 
    Notice that these charges only involve three degrees of freedom in the stress tensor, while two other degrees of freedoms, $T^1_{~1}$ and $T^1_{~2}$, are not used. Thus $L_n, \bar{L}_{\bar{n}}, M_{n,\bar{n}}$ are not all the modes in the expansion of the stress tensor. Another remarkable thing  is that the expressions of BMS$_4$ charges in terms of $\phi(\vec{k})$ and $\pi(\vec{k})$ modes are not as simple as the ones in 2D case. Actually, the straightforward generalization of the expressions of BMS$_3$ charges \eqref{eq:2DBMSGeneratorsInModes} in this case should be $\int d^2\vec{k} ~ f(\vec{k}) \pi(\vec{k})\partial^n \phi(\vec{k})$. However, one can check that the explicit expressions of BMS$_4$ charges are much more complicated than this expression, especially for negative indices. 
    This is due to the power-law factors in \eqref{eq:BMS4GeneratorsInStressTensor}, and the essential reason underlying is that the Fourier bases of spacial $\mathbb{R}^2$ are not closed under extended BMS$_4$ transformations.\par

\subsection{Canonical quantization}\label{subsec:3DMagneticScalarCanonicalQuantization}

    In this subsection, we discuss the standard canonical quantization and match the correlation functions with those from path integral. It turns out that the result is the natural extension of the $2$-dimensional case studied in section \ref{subsec:2DMagneticScalarCanonicalQuantization}. \par

\subsubsection{Canonical quantization and the Hilbert space}  \label{subsubsec:CanonicalQuantizationOf3DMagneticScalar}

    The Hilbert space is defined on the equal-time slice. Since the fields are real, the Hermitian conjugation are defined as
    \begin{equation}
        \begin{aligned}
            &\phi^\dagger(x) = \phi(x), && 
            \pi^\dagger(x) = \pi(x), &&
            &\phi^\dagger(\vec{k}) = \phi(-\vec{k}), && 
            \pi^\dagger(\vec{k}) = \pi(-\vec{k}). \\ 
        \end{aligned}
    \end{equation}
    The canonical momentum of field $\phi$ is $\Pi_\phi = -\pi$, thus the commutation relation is defined as usual:
    \begin{equation}\label{eq:3dMagneticScalarCanonitcalCommutationRelation}
        [\phi(t, \vec{x}_1), -\pi(t, \vec{x}_2)] = i\delta^{(2)}(\vec{x}_1-\vec{x}_2), \qquad [\phi(\vec{k}), \pi(\vec{p})] = \frac{-i}{4\pi^2}\delta^{(2)}(\vec{k}+\vec{p}),
    \end{equation} 
    and other commutations are vanishing. Here $\delta^{(2)}(\vec{x}) = \delta(x^1)\delta(x^2)$ is the delta-function of the $2$-dimensional spatial coordinates, and $\delta^{(2)}(\vec{k})$ is similar. The micro-causality is immediately read by plugging in the commutation relations:
    \begin{equation}
        \begin{aligned}\relax 
            [\phi(x_1),\phi(x_2)]&=0, \\
            [\phi(x_1),\pi(x_2)]&=-i\delta^{(2)}(\vec{x}_1-\vec{x}_2), \\
            [\pi(x_1),\pi(x_2)]&= i(t_1-t_2)\partial_i\partial_i\delta^{(2)}(\vec{x}_1-\vec{x}_2). \\
        \end{aligned}
    \end{equation}
    The commutators vanish once the spatial coordinates are not equal to each other $\vec{x}_1\neq\vec{x}_2$, which means that the information stays at fixed spatial point. \par

    The Hamiltonian $H = iM_{0,0}$ in these modes is lower bounded:
    \begin{equation}
        \begin{aligned}
            H &= \frac{1}{2} \int d^2x ~ \vec{k}^2\phi(\vec{k})\phi(-\vec{k}) = \frac{1}{2} \int d^2x ~ \vec{k}^2\abs*{\phi(\vec{k})}^2 \geq 0.
        \end{aligned}
    \end{equation}
     Similar to the case in section \ref{sec:2DMagneticScalar}, we can reorganize the modes as pairs of Heisenberg algebras by Bogoliubov transformations:
    \begin{equation}
        \begin{aligned}
            &\left\{\begin{aligned}
                \phi^c(\vec{k}) &= \frac{1}{\sqrt{2}}(\phi(\vec{k}) + \phi(-\vec{k})),\\
                \phi^s(\vec{k}) &= \frac{-i}{\sqrt{2}}(\phi(\vec{k}) - \phi(-\vec{k})),\\
            \end{aligned}\right. \\[10pt]
            &\left\{\begin{aligned}
                \pi^c(\vec{k}) &= \frac{1}{\sqrt{2}}(\pi(\vec{k}) + \pi(-\vec{k})),\\
                \pi^s(\vec{k}) &= \frac{-i}{\sqrt{2}}(\pi(\vec{k}) - \pi(-\vec{k})),\\
            \end{aligned}\right. \qquad\qquad 
            \begin{aligned}
                & k_1\geq0, k_2\in\mathbb{R}.
            \end{aligned}
        \end{aligned}
    \end{equation}
    These modes are Hermitian, 
    \begin{equation}
        \begin{aligned}
            &(\phi^c(\vec{k}))^\dagger = \phi^c(\vec{k}), \qquad  (\phi^s(\vec{k}))^\dagger = \phi^s(\vec{k}), \\
            &(\pi^c(\vec{k}))^\dagger = \pi^c(\vec{k}), \qquad (\pi^s(\vec{k}))^\dagger = \pi^s(\vec{k}),
        \end{aligned}
    \end{equation}
    and they do form  pairs of the Heisenberg algebras
    \begin{equation}
        \{\phi^c(\vec{k}),\pi^c(\vec{k})\}, \qquad \{\phi^s(\vec{k}),\pi^s(\vec{k})\},
    \end{equation}
    with the commutation relations being
    \begin{equation}
        \begin{aligned}\relax 
            [\phi^c(\vec{k}_1),\pi^c(\vec{k}_2)]  = \frac{-i}{4\pi^2}\delta^{(2)}(\vec{k}_1-\vec{k}_2), \qquad
            [\phi^s(\vec{k}_1),\pi^s(\vec{k}_2)]  = \frac{-i}{4\pi^2}\delta^{(2)}(\vec{k}_1-\vec{k}_2). \\
        \end{aligned}
    \end{equation}
     The fundamental fields in terms of these modes are
    \begin{equation}
        \begin{aligned}
            \phi(x) &= \sqrt{2}\int_{0}^{+\infty} dk_1 \int_{-\infty}^{+\infty} dk_2 ~ \phi^c(\vec{k}) \cos{(\vec{k}\cdot \vec{x})} + \phi^s(\vec{k}) \sin{(\vec{k}\cdot \vec{x})} , \\
            \pi(x) &= \sqrt{2}\int_{0}^{+\infty} dk_1 \int_{-\infty}^{+\infty} dk_2 ~ \left(\pi^c(\vec{k}) + t \vec{k}^2~\phi^c(\vec{k})\right) \cos{(\vec{k}\cdot \vec{x})} \\
                & \qquad\qquad\qquad\qquad\qquad\qquad\qquad + \left(\pi^s(\vec{k}) + t \vec{k}^2~\phi^s(\vec{k})\right) \sin{(\vec{k}\cdot \vec{x})}. \\
        \end{aligned}
    \end{equation}
    \par

    The bases in the rigged Hilbert space are chosen to be the eigenstates of $\phi(\vec{k})$ modes, and thus the basis states are the eigenstates of the Hamiltonian. To be more specific, the basis states are defined as direct product of $\ket*{\alpha(\vec{k})}$ states,
    \begin{equation}
        \ket{\alpha} = \prod_{\vec{k}\in\mathbb{R}^2} \ket*{\alpha(\vec{k})},
    \end{equation}
    in which $\prod_{\vec{k}\in\mathbb{R}^2}$ denotes the generalized direct product with continuous parameter $\vec{k}\in\mathbb{R}^2$. The states $\ket*{\alpha(\vec{k})}$ are the eigenstates of $\phi(\vec{k})$ modes, therefore $\ket{\alpha}$ is an eigenstate of $\phi(\vec{k})$ modes as well
    \begin{equation}
        \phi(\vec{k})\ket{\alpha} = \prod_{\vec{k}^\prime\in\mathbb{R}^2} \phi(\vec{k})\ket*{\alpha(\vec{k}^\prime)} = \prod_{\vec{k}^\prime\in\mathbb{R}^2} \alpha(\vec{k})\ket*{\alpha(\vec{k}^\prime)} = \alpha(\vec{k})\ket{\alpha}.
    \end{equation}
    The vacuum is the lowest energy state,
    \begin{equation}
        \ket{\text{vac}} =\ket{\alpha = 0}.
    \end{equation} 
    The conjugation of the basis states are defined as
    \begin{equation}\label{eq:3Doutstate}
        \bra{\alpha} = \prod_{\vec{k}\in\mathbb{R}^2} \bra*{\alpha(\vec{k})}.
    \end{equation}
    These basis states are $\delta$-function-like normalized, 
    \begin{equation}\label{eq:3Dnormalization}
        \braket{\tilde{\alpha}}{\alpha} = \prod_{\vec{k},\vec{k}^\prime\in\mathbb{R}^2} \bra*{\tilde{\alpha}(\vec{k}^\prime)}\ket*{\alpha(\vec{k})} 
        = \prod_{\vec{k},\vec{k}^\prime\in\mathbb{R}^2} \delta(\alpha(\vec{k}) - \tilde{\alpha}(\vec{k}^\prime))\equiv \delta(\alpha -\tilde{\alpha}) .
    \end{equation}
    Here $\delta(\alpha -\tilde{\alpha})$ is defined as the product of delta functions $\delta(\alpha(\vec{k}) - \tilde{\alpha}(\vec{k}^\prime))$ for every $\vec{k},\vec{k}^\prime\in\mathbb{R}^2$. In fact, it is natural to view $\alpha(\vec{k})$ as a function of $\vec{k}$, and $\delta(\alpha - \tilde{\alpha})$ as the delta function in the sense of functional integral. For a functional $F(\alpha)$, we have
    \begin{equation}
        \int [\mathcal{D}\alpha] ~ \delta(\alpha -\tilde{\alpha}) F[\alpha] = F(\tilde{\alpha}).
    \end{equation}
    Similar to the case in $2$D, the vacuum expectation value of operator $\mathcal{O}$ is defined as
    \begin{equation}
        \expval{\mathcal{O}} = \lim_{\alpha\to0} \left(\lim_{\tilde{\alpha}\to\alpha} \frac{\bra{\tilde{\alpha}}\mathcal{O}\ket{\alpha} + \bra{\alpha}\mathcal{O}\ket{\tilde{\alpha}}}{2\bra{\tilde{\alpha}}\ket{\alpha}}\right) .
    \end{equation}
    \par

    We can also define the eigenstates of $\pi(\vec{k})$ modes as
    \begin{equation}
        \ket{\kappa} = \prod_{\vec{k}\in\mathbb{R}^2} \ket*{\kappa(\vec{k})}.
    \end{equation}
    The inner product of $\ket{\kappa}$ and $\bra{\alpha}$ is
    \begin{equation}
        \begin{aligned}
            &\braket{\alpha}{\kappa}  = \braket{\kappa}{\alpha}^* = \prod_{\vec{k}\in\mathbb{R}^2} \sqrt{2\pi} \exp{-4\pi^2 i(\kappa(\vec{k})\alpha(-\vec{k}))} . \\
                &\qquad\qquad\qquad\quad \propto \exp{-4\pi^2 i \int d^2k \left(\kappa(\vec{k})\alpha(-\vec{k})\right)},
        \end{aligned}
    \end{equation}
    and the identity operator is defined as $\mathbb{I} = \int [\mathcal{D}\alpha] \ketbra{\alpha}$ or $\mathbb{I} = \int [\mathcal{D}\kappa] \ketbra{\kappa}$. Using the identity operator, it can be checked that the eigenstates of $\phi(\vec{k})$ modes are generated by an operator $V[\alpha]$ defined on the $t=0$ surface,
    \begin{equation}
        V[\alpha] = N\exp{- i \int d^2 k ~ \alpha(\vec{k}) \pi(t=0,\vec{x})e^{-i\vec{k}\cdot\vec{x}}}.
    \end{equation}
    The fact that the operator $V[\alpha]$ is a $2$-dimensional operator indicates that there is no well-defined state-operator correspondence in $3$-dimensional magnet scalar theory. \par

    Similar to the calculation in the 2D case, the 2-point correlators of the fundamental fields are 
    \begin{equation}
        \begin{aligned}
            &\expval{\phi(x_1)\phi(x_2)} = 0, \\
            &\expval{\phi(x_1)\pi(x_2)} = \frac{-i}{2} \delta^{(2)}(\vec{x}_{12}), \\[5pt]
            &\expval{\pi(x_1)\phi(x_2)} = \frac{i}{2} \delta^{(2)}(\vec{x}_{12}), \\[5pt]
            &\expval{\pi(x_1)\pi(x_2)} = \frac{1}{\epsilon^2}f(\vec{x}_{12}) + \frac{it_{12}}{2} \vec{\partial}^2\delta^{(2)}(\vec{x}_{12}) .\\
        \end{aligned}
    \end{equation}
    Here $f(\vec{x})$ is a generic function, whose form relies on the explicit regularization  scheme, and it satisfies $f^*(\vec{x}) = f(-\vec{x})$. It is straightforward to get the time-ordered propagators:
    \begin{equation}\label{eq:FeynmanPropagatorOf3DMagneticScalar}
        \begin{aligned}
            &\expval{\mathcal{T}\left\{\phi(x_1)\phi(x_2)\right\}} = 0, \\
            &\expval{\mathcal{T}\left\{\phi(x_1)\pi(x_2)\right\}} = \frac{-i}{2} \text{sign}(t_{12}) \delta^{(2)}(\vec{x}_{12}),\\[5pt]
            &\expval{\mathcal{T}\left\{\pi(x_1)\phi(x_2)\right\}} = \frac{i}{2} \text{sign}(t_{12}) \delta^{(2)}(\vec{x}_{12}),\\[5pt]
            &\expval{\mathcal{T}\left\{\pi(x_1)\pi(x_2)\right\}} = \frac{1}{\epsilon^2}(\theta(t_{12})f(\vec{x}_{12})+\theta(-t_{12})f(-\vec{x}_{12})) + \frac{i\abs{t_{12}}}{2} \vec{\partial}^2\delta^{(2)}(\vec{x}_{12}) .\\
        \end{aligned}
    \end{equation}
    As will be shown below, they agree with the ones calculated from the path integral. \par

    At last, we briefly discuss the quantum BMS$_4$ symmetry. In the quantization, we use the Weyl ordering as the normal ordering. For example, we have
    \begin{equation}
        \pi\partial_1\phi \longrightarrow \frac{1}{2}\pi\partial_1\phi + \frac{1}{2}\partial_1\phi \pi.
    \end{equation}
    Plugging this into the stress tensor, we find that the VEV of the stress tensor is vanishing
    \begin{equation}
        \expval{T^\mu_{~\nu}} = 0.
    \end{equation}
    Hence the generators of the BMS$_4$ symmetry have vanishing VEVs,
    \begin{equation}
        \expval{L_{n}} = \expval{\bar{L}_{\bar{n}}} = \expval{M_{n,\bar{n}}}.
    \end{equation}
    Similar to the $2$D case, we find that the BMS$_4$ algebra is anomaly free and the commutation relations are
    \begin{equation}\label{eq:3dcomm}
        \begin{aligned}
            &[L_m,L_n] = (m-n) L_{m+n}, \quad [\bar{L}_{\bar{m}},\bar{L}_{\bar{n}}] = (\bar{m}-\bar{n}) \bar{L}_{\bar{m}+\bar{n}}, \quad [L_{m},\bar{L}_{\bar{m}}] = 0, \\
            &[L_m,M_{n,\bar{n}}] = \left(\frac{m}{2} - n \right)M_{m+n,\bar{n}}, \quad [\bar{L}_{\bar{m}},M_{n,\bar{n}}] = \left(\frac{\bar{m}}{2} - \bar{n} \right)M_{n,\bar{m}+\bar{n}}.\\
        \end{aligned} 
    \end{equation}
    Using the canonical commutation relations of the fundamental fields \eqref{eq:3dMagneticScalarCanonitcalCommutationRelation}, it can be checked that the actions of BMS$_4$ generators on the fields agree with \eqref{eq:3DInfinitesimalBMS4Transformation}. \par

\subsubsection{Path integral quantization}

    The calculations in path integral are similar to the ones in $2$D. We first symmetrize the action \eqref{eq:Action3DMagneticScalar},
    \begin{equation}
        \begin{aligned}
            S & = - \frac{1}{2}\int d^3x ~ 2\pi \partial_t \phi + \partial_i \phi \partial_i \phi + \partial_t(-\pi\phi) + \partial_i(-\phi\partial_i\phi) \\
                & = \int d^3x ~ \frac{1}{2} \phi \partial_t \pi -\frac{1}{2} \pi \partial_t \phi + \frac{1}{2} \phi \vec{\partial}^2  \phi\\
                & = \int d^3x ~ \bm{\Phi}^\dagger\hat{D}\bm{\Phi} , \\
        \end{aligned}
    \end{equation}
    in which the derivative operator and the fundamental fields are denoted by
    \begin{equation}
        \hat{D} = \begin{pmatrix}\frac{1}{2}\vec{\partial}^2 & \frac{1}{2}\partial_t \\  -\frac{1}{2}\partial_t & 0\end{pmatrix}, \qquad
        \bm{\Phi} = \begin{pmatrix}\phi \\ \pi \end{pmatrix}.
    \end{equation}
    Therefore, the sourced partition function for this theory is 
    \begin{equation}
        \begin{aligned}
            \mathcal{Z}[\bm{J}] & = \int \mathcal{D}\phi\mathcal{D}\pi ~ \exp{iS + i\int d^3x ~ J_\phi\phi + J_\pi\pi } \\
                & = \int \mathcal{D}\phi\mathcal{D}\pi ~ \exp{i \int d^3x ~ \bm{\Phi}^\dagger\hat{D}\bm{\Phi} + \frac{1}{2} \bm{J}^\dagger\bm{\Phi} + \frac{1}{2} \bm{\Phi}^\dagger \bm{J} } \\
                & = N \exp{i \int d^3x_1 d^3x_2 -\frac{1}{4} \bm{J}^\dagger(x_1)\hat{D}^{-1}(x_1-x_2) \bm{J}(x_2)} . \\
        \end{aligned}
    \end{equation}
    Here $N$ is the overall factor, and $G(x_1-x_2) = \hat{D}^{-1}(x_1-x_2)$ is the Green's function respecting to the derivative operator $\hat{D}$. After Fourier transformation, we have
    \begin{equation}\label{eq:GreenFunctionOf3DMagneticScalar}
        \begin{aligned}
            G(x) &= \int \frac{d\omega d^2k}{(2\pi)^3} \begin{pmatrix} -\frac{\vec{k}^2}{2} & -\frac{i\omega}{2}  \\ \frac{i \omega }{2} & 0 \\ \end{pmatrix}^{-1} e^{-i\omega t - i kx} 
                = \int \frac{d\omega d^2k}{(2\pi)^3} \begin{pmatrix} 0 & -\frac{2 i}{\omega } \\ \frac{2 i}{\omega } & \frac{2 \vec{k}^2}{\omega ^2} \\\end{pmatrix}  e^{-i\omega t - i kx} \\[10pt]
                &= \begin{pmatrix} 
                    0 & -\text{sign}(t)\delta^{(2)}(\vec{x}) \\ 
                    \text{sign}(t)\delta^{(2)}(\vec{x}) & A(\vec{x}) + \abs{t}\vec{\partial}^2\delta^{(2)}(\vec{x}) \\ 
                \end{pmatrix}.
        \end{aligned}
    \end{equation}
    The function $A(\vec{x})$ in the $G_{\pi\pi}$ component is a power-law divergent term. As in the $2$D case, the divergence in this term comes from the integral of $\omega$, and the explicit form of $A(\vec{x})$ depends on the regularization scheme. We can further get the propagators of the fundamental fields, which matches exactly with the ones read in the  canonical quantization \eqref{eq:FeynmanPropagatorOf3DMagneticScalar},
    \begin{equation}
        \begin{aligned}
            &\expval{\bm{\Phi}_i(x_1)\bm{\Phi}_j(x_2)} = \frac{i}{4}(G_{ij}(x_{12}) + G_{ji}(x_{21}))  \\
                & \qquad = \begin{pmatrix} 
                    0 & -\frac{i}{2}\text{sign}(t_{12})\delta^{(2)}(\vec{x}_{12}) \\ 
                    \frac{i}{2}\text{sign}(t_{12})\delta^{(2)}(\vec{x}_{12}) & i\abs{t_{12}}\vec{\partial}^2\delta^{(2)}(\vec{x}_{12}) + A(\vec{x}_{12}) \\ 
                \end{pmatrix}. \\
        \end{aligned}
    \end{equation}

\subsection{Non-unitary canonical quantization}\label{subsec:3DHighestWeightQuantization}

    As discussed earlier, the BMS$_4$ generators have complicated expressions in terms of $\phi(\vec{k})$ and $\pi(\vec{k})$ modes. Therefore, it is hard to read the action of the modes on the vacuum from the highest-weight condition, i.e. the positive BMS$_4$ charges annihilate the vacuum. In this subsection, we directly define the highest-weight vacuum by the action of the modes, and discuss the form of the correlation functions.\par

\begin{figure}[htbp]
        \centering
        (a)~~~~~\includegraphics[width=8.6cm,align=c]{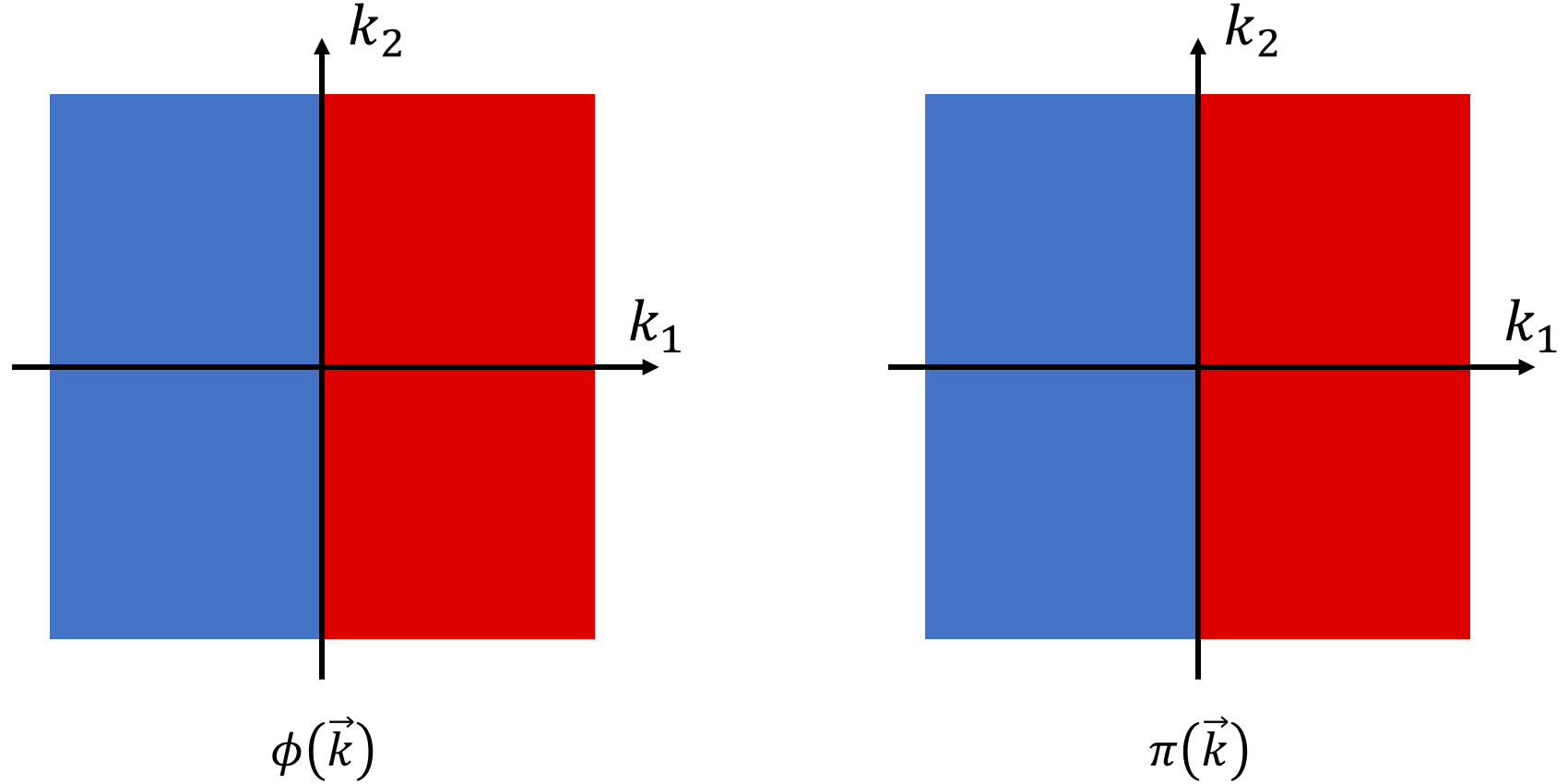}
        \\[5pt]
        (b)~~~~~\includegraphics[width=8.6cm,align=c]{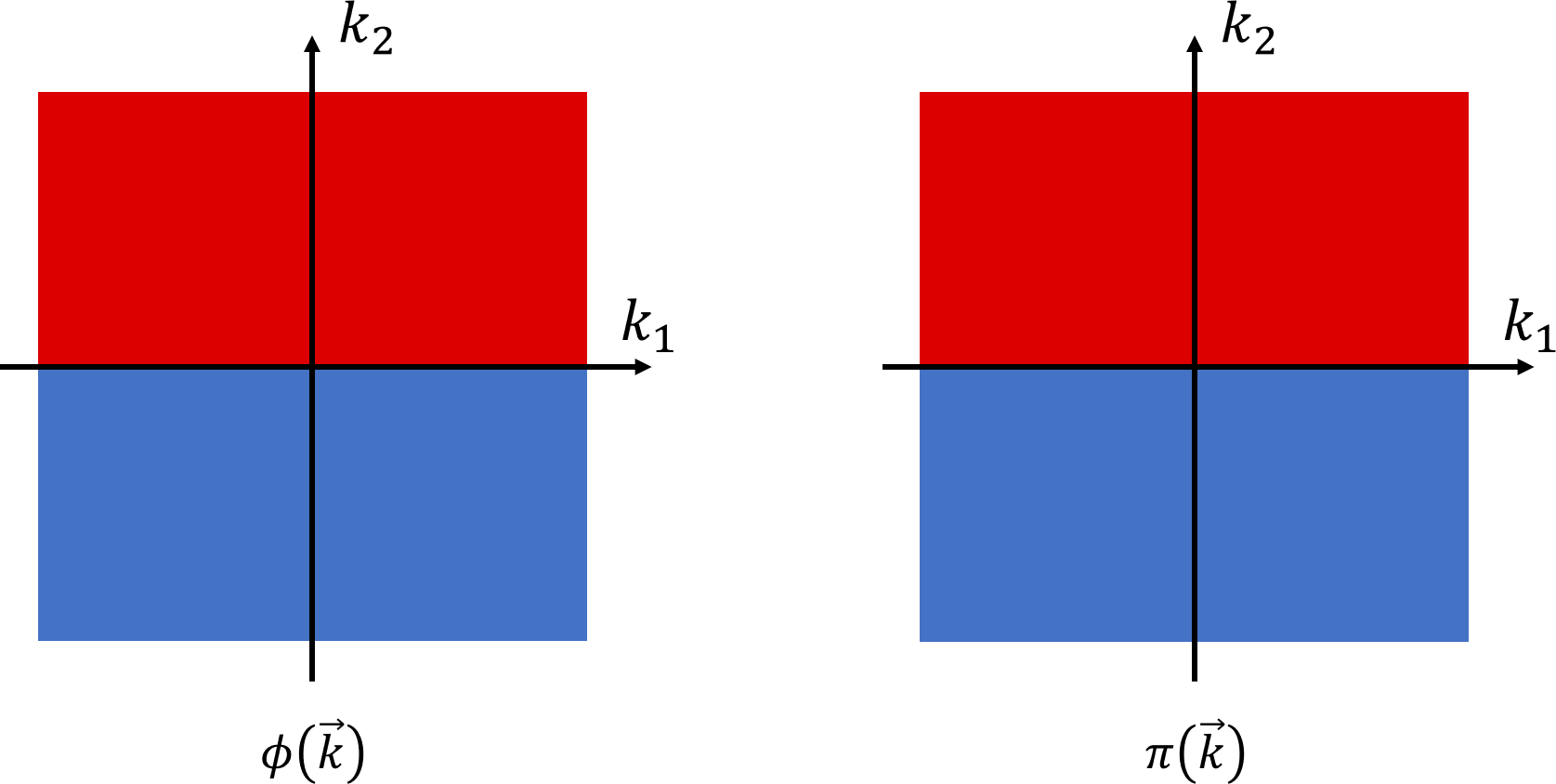}
        \\[5pt]
        (c)~~~~~\includegraphics[width=8.6cm,align=c]{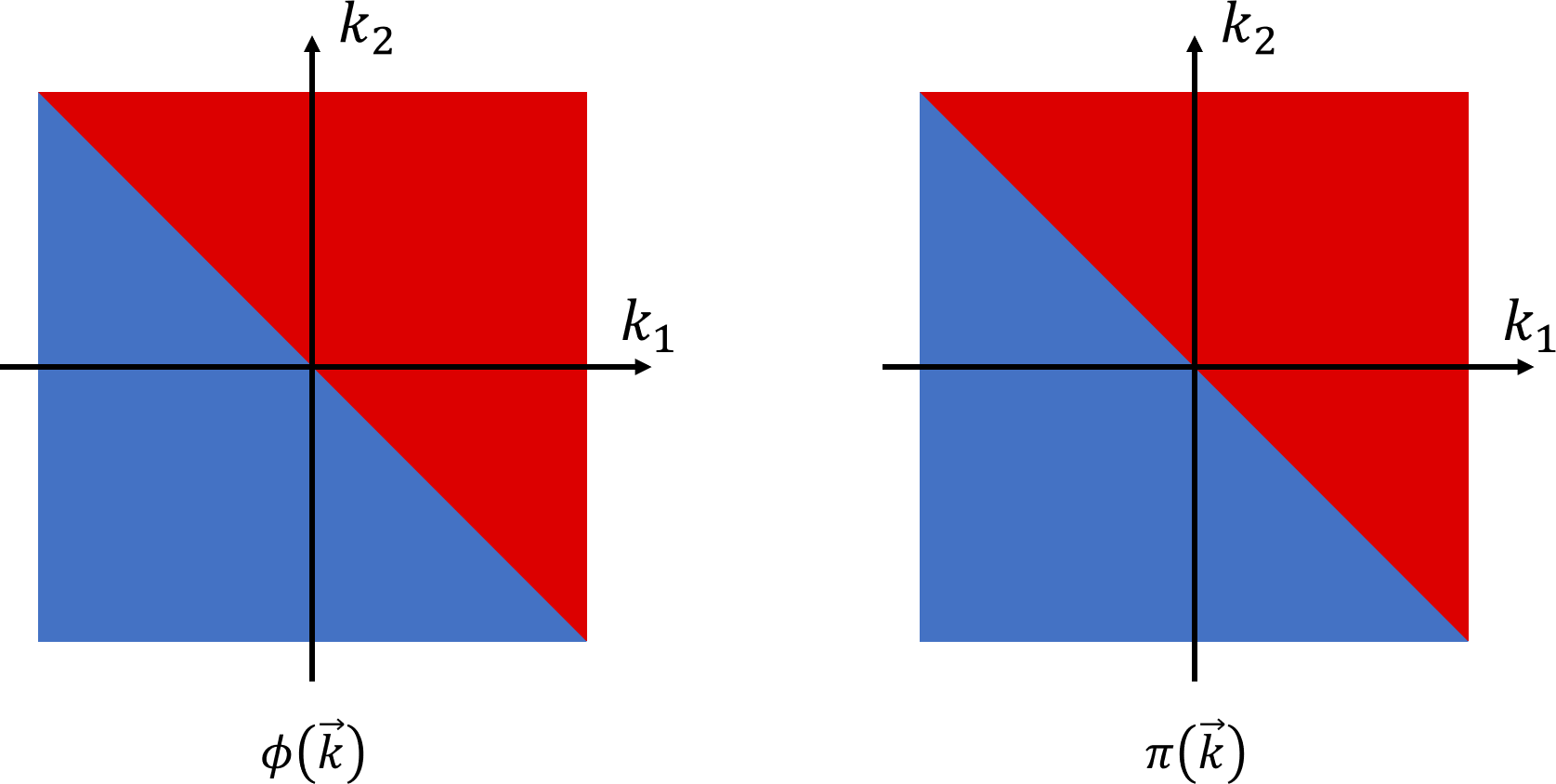}
        \captionsetup{margin = 3em}
        \caption{\centering  (a–c) provides three definitions of the highest-weight vacuum in 3D magnetic scalar theory. In each case, the red parts annihilate the in-vacuum $\ket{\text{vac}}_h$, while the blue parts annihilate the out-vacuum $_h\!\!\bra{\text{vac}}$.} 
        \label{fig:RxR2HighestWeight}
    \end{figure}

    In order to keep the canonical commutation relations between $\phi(\vec{k})$ and $\pi(\vec{k})$, if $\phi(\vec{k})$ annihilates the in-vacuum $\phi(\vec{k})\ket{\text{vac}}_h = 0$ for a given $\vec{k}$, then $\pi(-\vec{k})$ should not annihilate the in-vacuum $\pi(-\vec{k})\ket{\text{vac}}_h \neq 0$, and at the same time, the out-vacuum satisfies $_h\!\!\bra{\text{vac}}\phi(\vec{k}) \neq 0$ and $_h\!\!\bra{\text{vac}}\pi(-\vec{k}) = 0$. This statement remains true after exchanging $\phi$ and $\pi$. Hence, one definition of the highest-weight vacuum is
    \begin{equation}\label{eq:3DHighestWeightVacuum2}
        \begin{aligned}
            \phi(\vec{k})\ket{\text{vac}}_h &= 0, & \pi(\vec{k})\ket{\text{vac}}_h &= 0, & k_1 > 0, \\
            _h\!\!\bra{\text{vac}}\phi(\vec{k}) &= 0, &_h\!\!\bra{\text{vac}}\pi(\vec{k}) &= 0, & k_1<0,\\
        \end{aligned}
    \end{equation}
    which is illustrated in Figure \ref{fig:RxR2HighestWeight}(a). The correlation functions of the fundamental fields are
    \begin{equation}
        \begin{aligned}
            &\expval{\phi(x_1)\phi(x_2)} = 0, \\
            &\expval{\phi(x_1)\pi(x_2)} = \frac{-1}{2\pi}\frac{1}{x^1_{12}}\delta(x^2_{12}), \\[5pt]
            &\expval{\pi(x_1)\phi(x_2)} = \frac{1}{2\pi}\frac{1}{x^1_{12}}\delta(x^2_{12}), \\[5pt]
            &\expval{\pi(x_1)\pi(x_2)} = \frac{-t_{12}}{2\pi} \left(\frac{2}{(x^1_{12})^3}\delta(x^2_{12}) + \frac{1}{x^1_{12}}\partial^2\delta(x^2_{12})\right).\\
        \end{aligned}
    \end{equation}
    It is immediately noticed that, the correlation functions are power-law in $x^1$ and $\delta$-functions in $x^2$. This form breaks the rotational symmetry, therefore the above definition of the highest-weight vacuum seems not to be physically meaningful. Moreover, there are two other options in defining the highest-weight vacuum, as shown in \ref{fig:RxR2HighestWeight}(b) and \ref{fig:RxR2HighestWeight}(c). Neither of them is physically acceptable, as they also break the rotational symmetry.

    To conclude this subsection, we find that in the highest-weight quantization schemes, there is no satisfactory  definition of the vacuum such that the corresponding correlation functions are of power-law forms in the spatial directions as well as of the rotational symmetry. \par

\section{Electric scalar in \texorpdfstring{$\mathbb{R}\times\mathbb{R}^2$}{R×R2}} \label{sec:3DElectricScalar}

    The quantization of the $d=3$ electric scalar theory with BMS$_4$ symmetry is also straightforward. The BMS$_4$ invariant electric scalar theory in \texorpdfstring{$\mathbb{R}\times\mathbb{R}^2$}{R×R2} is given by
    \begin{equation}\label{eq:Action3DElectricScalar}
        S = \int d^3 x ~ \frac{1}{2}(\partial_t \phi)^2 .
    \end{equation}
    The equations of motion of the fundamental fields can be read as
    \begin{equation}
        \phi: ~ \partial_t^2 \phi = 0.     
    \end{equation}
    The fields can be expanded as
    \begin{equation}
        \phi(x) = \int d^2k ~ (\phi(\vec{k}) + t \chi(\vec{k})) e^{-i \vec{k}\cdot\vec{x}}.
    \end{equation} 
    To see the manifest BMS$_4$ symmetry of the action, recall that the field $\phi$ is a BMS$_4$ scalar and it transforms under BMS$_4$ transformations \eqref{eq:BMS4Transformation} as
    \begin{equation}
        \Tilde{\phi}(\Tilde{x}) = (\partial F \bar{\partial}\bar{F})^{-\frac{1}{4}}\phi(x),
    \end{equation}
    and infinitesimally
    \begin{equation}\label{eq:BMS4scalartransform}
            \delta \phi(x) = \Tilde{\phi}(\tilde{x}) - \phi(x) = -\frac{1}{2}\partial_t\xi^0\phi.
    \end{equation}
    
    Given in \cite{Dutta:2022vkg}, the stress tensor for $3$D electric scalar theory is
    \begin{equation}\label{eq:ElectricT}
        T^\mu_{~\nu} = 
	\begin{pmatrix}
		\frac{1}{2}(\partial_t \phi)^2&\frac{3}{4}\partial_t \phi \partial_1 \phi-\frac{1}{4} \phi \partial_t\partial_1 \phi&\frac{3}{4}\partial_t \phi \partial_2 \phi-\frac{1}{4} \phi \partial_t\partial_2 \phi\\
		0&-\frac{1}{4}(\partial_t \phi)^2&0\\
		0&0&-\frac{1}{4}(\partial_t \phi)^2\\
	\end{pmatrix}.
    \end{equation}
    This stress tensor satisfies the conserved current equation and is chosen to be traceless. The BMS$_4$ charges are formally the same as the magnetic sector, with the stress tensor replaced by the electric one \eqref{eq:ElectricT},
    \begin{equation}\label{eq:ElectricBMS4GeneratorsInStressTensor}
        \begin{aligned}
            &L_n = i\int d^2x ~ \frac{n+1}{2}t(x^1+ix^2)^n T^0_{~0} + \frac{1}{2}(x^1+ix^2)^{(n+1)} (T^0_{~1} - iT^0_{~2}), \\
            &\bar{L}_{\bar{n}} = i\int d^2x ~ \frac{\bar{n}+1}{2}t(x^1-ix^2)^{\bar{n}} T^0_{~0} + \frac{1}{2}(x^1+ix^2)^{(\bar{n}+1)} (T^0_{~1} + iT^0_{~2}), \\
            &M_{n,\bar{n}} = i \int d^2x ~ (x^1+ix^2)^n (x^1-ix^2)^{\bar{n}} T^0_{~0} . \\
        \end{aligned}
    \end{equation}

    \subsection{Canonical quantization and the Hilbert space}\label{subsec:3DElectricScalarCanonicalQuantization}
    In this subsection, we discuss the standard canonical quantization and calculate the correlation functions of the electric BMS$_4$ scalar theory. Again we perform the canonical quantization on the equal-time slice. The Hermitian conjugate conditions are
    \begin{equation}
        \phi^\dagger(x) = \phi(x) ~
        \Longleftrightarrow  ~
        \phi^\dagger(\vec{k}) = \phi(-\vec{k}), ~
        \chi^\dagger(\vec{k}) = \chi(-\vec{k}).
    \end{equation}
Since the canonical conjugate of the field $\phi$ is $\Pi_\phi = \partial_t \phi$, the canonical commutation relation is
    \begin{equation}
        [\phi(x),\partial_t \phi(x)] = i\delta^{(2)}(\vec x_1-\vec x_2) ~\Longleftrightarrow~ [\phi(\vec k),\chi(\vec p)] = \frac{i}{4\pi^2} \delta^{(2)}(\vec{k}+\vec{p}).
    \end{equation}
    Furthermore, we can also perform a Bogoliubov transformations to organize the modes into Heisenberg pairs $\{\phi^c(\vec{k}),\chi^c(\vec{k})\}, ~ \{\phi^s(\vec{k}),\chi^s(\vec{k})\}$:
    \begin{equation}
    \begin{aligned}
        &\left\{\begin{aligned}
            \phi^c(\vec k) &= \frac{1}{\sqrt{2}}(\phi(\vec k) + \phi(-\vec k))\\
            \phi^s(\vec k) &= \frac{-i}{\sqrt{2}}(\phi(\vec k) - \phi(-\vec k)),\\
        \end{aligned}\right. \\
        &\left\{\begin{aligned}
            \chi^c(\vec k) &= \frac{1}{\sqrt{2}}(\chi(\vec k) + \chi(-\vec k))\\
            \chi^s(\vec k) &= \frac{-i}{\sqrt{2}}(\chi(\vec k) - \chi(-\vec k))\\
        \end{aligned}\right. \qquad\qquad
        \begin{aligned}
        & k_1\geq 0, k_2\in\mathbb{R},
        \end{aligned}
    \end{aligned}
    \end{equation}
    with 
    \begin{equation}
        \begin{aligned}
            &(\phi^c(\vec{k}))^\dagger = \phi^c(\vec{k}), \qquad  (\phi^s(\vec{k}))^\dagger = \phi^s(\vec{k}), \\
            &(\chi^c(\vec{k}))^\dagger = \chi^c(\vec{k}), \qquad (\chi^s(\vec{k}))^\dagger = \chi^s(\vec{k}).
        \end{aligned}
    \end{equation} The commutation relations are realized as 
    \begin{equation}
        \begin{aligned}\relax 
            [\phi^c(\vec{k}_1),\chi^c(\vec{k}_2)]  = \frac{i}{4\pi^2}\delta^{(2)}(\vec{k}_1-\vec{k}_2), \qquad
            [\phi^s(\vec{k}_1),\chi^s(\vec{k}_2)]  = \frac{i}{4\pi^2}\delta^{(2)}(\vec{k}_1-\vec{k}_2). \\
        \end{aligned}
    \end{equation}

    Expanding in these modes, the scalar field can be expressed by
    \begin{equation}
    \begin{aligned}
        \phi(x) &= \sqrt{2}\int_{0}^{+\infty} dk_1 \int_{-\infty}^{+\infty} dk_2 ~ \left(\phi^c(\vec{k}) + t ~\chi^c(\vec{k})\right) \cos{(\vec{k}\cdot \vec{x})} \\
            & \qquad\qquad\qquad\qquad\qquad\qquad\qquad + \left(\phi^s(\vec{k}) + t ~\chi^s(\vec{k})\right) \sin{(\vec{k}\cdot \vec{x})}. \\
    \end{aligned}
    \end{equation} The Hamiltonian $H = i M_{0,0}$ in these modes is clearly lower bounded
    \begin{equation}
    \begin{aligned}
                    H  &= \frac{1}{2} \int d^2 k ~ \chi(\vec{k})\chi(-\vec{k}) = \frac{1}{2} \int d^2 k ~ \abs*{\chi(\vec{k})}^2\\
                    &= \frac{1}{4} \int d^2 k ~  \left(\chi^c(\vec{k})^2 + \chi^s(\vec{k})^2\right)\geq 0. 
    \end{aligned}
    \end{equation} Thus the Hilbert space is spanned by the eigenstates of the Hamiltonian $\ket{\alpha} = \prod_{\vec{k}\in\mathbb{R}^2} \ket*{\alpha(\vec{k})}$ consisting of direct products of Heisenberg modes $\chi(\vec{k})\ket*{\alpha(\vec{k}^\prime)} = \alpha(\vec{k})\ket*{\alpha(\vec{k}^\prime)}$ :
    \begin{equation}
        \chi(\vec{k})\ket{\alpha} = \prod_{\vec{k}^\prime\in\mathbb{R}^2} \chi(\vec{k})\ket*{\alpha(\vec{k}^\prime)} = \prod_{\vec{k}^\prime\in\mathbb{R}^2} \alpha(\vec{k})\ket*{\alpha(\vec{k}^\prime)} = \alpha(\vec{k})\ket{\alpha}.
    \end{equation}
    The induced vacuum is the lowest-energy state 
    \begin{equation}
        \ket{\text{vac}} =\ket{\alpha = 0} .
    \end{equation}
     One can construct the out-states as in \eqref{eq:3Doutstate},\eqref{eq:3Dnormalization} and obtain a $\delta$-function normalization
    \begin{equation}
        \braket{\tilde{\alpha}}{\alpha} = \prod_{\vec{k},\vec{k}^\prime\in\mathbb{R}^2} \bra*{\tilde{\alpha}(\vec{k}^\prime)}\ket*{\alpha(\vec{k})} 
        = \prod_{\vec{k},\vec{k}^\prime\in\mathbb{R}^2} \delta(\alpha(\vec{k}) - \tilde{\alpha}(\vec{k}^\prime))\equiv \delta(\alpha -\tilde{\alpha}) .
    \end{equation}
    Similarly, the eigenstates $\ket{\alpha}$ are generated by the surface operators $V[\alpha]$, 
    \begin{equation}
        V[\alpha] = N\exp{ i \int d^2 k ~ \alpha(\vec{k}) \phi(t=0,\vec{x})e^{-i\vec{k}\cdot\vec{x}}}.
    \end{equation} Upon subtracting the renormalization factor, the 2-point correlator is
    \begin{equation}\label{eq:2ptElectricScalar}
        \expval{\phi(x_1)\phi(x_2)} = \frac{it_{12}}{2} \delta^{(2)}(\vec{x}_{12}) ,
    \end{equation}
    and the time-ordered propagator is
    \begin{equation}
        \expval{\mathcal{T}\left\{\phi(x_1)\phi(x_2)\right\}} =  \frac{i\abs{t_{12}}}{2} \delta^{(2)}(\vec{x}_{12}),
    \end{equation} which again matches the result from path integral\cite{Chen:2023pqf}.

    For the electric theory, by using Weyl ordering the VEVs of BMS$_4$ symmetry generators are still vanishing,
    \begin{equation}
        \expval{L_{n}} = \expval{\bar{L}_{\bar{n}}} = \expval{M_{n,\bar{n}}}=0.
    \end{equation} The electric theory is also free of central charges and the quantum commutation relations are the same as \eqref{eq:3dcomm}, and the commutators of the BMS$_4$ operators with the scalar field $\phi(x)$ agree with \eqref{eq:BMS4scalartransform}. \par

\subsection{Non-unitary quantization}

    For the $3$D electric theory, the discussions of non-unitary quantization is in the same manner with magnetic case. Half of the $\phi(\vec k), \chi(\vec k)$ modes can be chosen to annihilate the vacuum in different quantization schemes. To be precise, for example, if $\chi(\vec{k})$ annihilates the in-vacuum $\chi(\vec{k})\ket{\text{vac}}_h = 0$ for a given $\vec{k}$, then $\phi(-\vec{k})$ can not annihilate it, and the out-vacuum satisfies $_h\!\!\bra{\text{vac}}\chi(\vec{k}) \neq 0, ~ _h\!\!\bra{\text{vac}}\phi(-\vec{k}) = 0$. Similar arguments hold if we exchange $\phi$ and $\chi$. We still have different choices of the highest-weight vacuum as in the magnetic theory:  for example, one of them is given by
    \begin{equation}
        \begin{aligned}
            \chi(\vec{k})\ket{\text{vac}}_h &= 0, & \phi(\vec{k})\ket{\text{vac}}_h &= 0, & k_1 > 0, \\
            _h\!\!\bra{\text{vac}}\phi(\vec{k}) &= 0, &_h\!\!\bra{\text{vac}}\chi(\vec{k}) &= 0, & k_1<0.\\
        \end{aligned}
    \end{equation} The correlation function of the fundamental fields in this specific highest-weight vacuum is
    \begin{equation}
        \expval{\phi(x_1)\phi(x_2)} =\frac{t_{12}}{x^1_{12}}\delta(x^2_{12}).\\
    \end{equation} This form again breaks the rotational symmetry. Moreover, it can be easily checked that there is no choice of the highest-weight vacuum without breaking the rotational symmetry.

\section{Discussion}\label{sec:Discussion}

    In this work, we discussed the quantizations of $d=2$ and $d=3$ Carrollian scalar theories, including  magnetic scalar theories in 2D and 3D and electric scalar theory in 3D. We realized the BMS symmetries in these theories and discussed two different quantization schemes. The standard canonical quantization yields unitary Hilbert space and the induced vacuum in the literature. In the induced vacuum,  the correlation functions exhibit the structure of a power-law form in the time direction and derivatives of Dirac delta-function in the spatial directions, $t^m\partial^n\delta^{(d-1)}$, and they are identical to the ones computed by using the path-integral quantization. It is worth mentioning that this quantization scheme is anomaly-free by definition. The other quantization scheme acquires a highest-weight vacuum and  sacrifice of the unitarity of the Hilbert space. In the $2$-dimensional case, the corresponding correlation functions are of power-law forms in the space-time coordinates, and they match the results from taking the $c\to0$ limit of CFT correlation functions. This quantization scheme is anomalous, and the anomaly has similar form with the one in $2$D CFT. However for the $3$-dimensional case, there is no physically meaningful highest-weight quantization scheme. \par

    In an earlier paper \cite{deBoer:2023fnj}, the authors have discussed the canonical quantization of massive Carrollian scalar theories on the plane $\mathbb{R}^{(d+1)}$. In the electric scalar case, the Hamiltonian have similar form with Lorentzian scalar theory:
    \begin{equation}
        H\propto \int d^{(d-1)} k ~ a^\dagger_{\vec{k}}a_{\vec{k}}.
    \end{equation}
     This indicates that the spectrum of the theory forms a ladder representation, and the vacuum is the lowest-energy state. The authors also calculated the 2-point correlation function:
    \begin{equation}
        \expval{\phi(t_1,\vec{x}_1)\phi(t_2,\vec{x}_2)}\propto e^{-im(t_1-t_2)}\delta^{(d-1)}(\vec{x}_1-\vec{x}_2).
    \end{equation}
    The  $m\to0$ limit of the correlation function is
    \begin{equation}
        \expval{\phi(t_1,\vec{x}_1)\phi(t_2,\vec{x}_2)}\propto \abs{t_1-t_2}\delta^{(d-1)}(\vec{x}_1-\vec{x}_2),
    \end{equation}
    which matches \eqref{eq:2ptElectricScalar} in canonical quantization. However, the discussion on the quantization is subtler in our case. The Hamiltonian of the massless scalar takes the form of 
    \begin{equation}\label{eq:GeneralHamiltonianForm}
        H  \propto \int d^2 k ~ \abs*{\phi(\vec{k})}^2,
    \end{equation}
    where $\phi(\vec{k})$ are commuting modes. Hence the definition of the vacuum should be treated with more care. Moreover, besides the induced vacuum which corresponds to the $m\to 0$ limit of \cite{deBoer:2023fnj}, we also discussed the possibility in choosing the highest-weight vacuum.

In the discussion of the canonical vacuum, we have introduced the rigged Hilbert space. The rigged Hilbert space is a triplet $\Phi\subseteq \mathcal{H}\subseteq \Phi^\times$, where $\mathcal{H}$ is the traditional Hilbert space, $\Phi$ is the space of the physical states, and $\Phi^\times$ is its dual. Considering $1$D quantum mechanics for an example, $\mathcal{H}$ is the space of square integrable functions, $\Phi$ is the space of rapidly decreasing functions, and $\Phi^\times$ is the space of tempered distributions. Due to the fact that the generic Hamiltonian of the massless scalar theories have the form of \eqref{eq:GeneralHamiltonianForm}, the energy eigenstates are non-normalizable states in $\Phi^\times$. By the help of the rigged Hilbert space, we can discuss the canonical quantization of the Carrollian massless scalar theories.       \par

    One remarkable thing we found is that the discussions of state-operator correspondence in CFT can not be directly extended to the Carrollian case. From the symmetry perspective, there is no BMS transformation that maps the past-infinity time slice of Carrollian manifold to a point. This means that a state in the Hilbert space can not correspond to a local operator. For example, as shown in section \ref{subsec:HilbertSpace}, the basis states of the canonical Hilbert space correspond to line operators rather than local vertex operators.  \par

\section*{Acknowledgments}

    We are grateful to Zhe-fei Yu, Reiko Liu, Pengxiang Hao, Zezhou Hu, Hongjie Chen for their valuable discussions and suggestions. This research is supported by NSFC Grant  No. 11735001, 12275004.

\appendix
\renewcommand{\appendixname}{Appendix~\Alph{section}}

\bibliographystyle{utphys}
\bibliography{refs.bib}
\end{document}